\documentclass[fleqn,usenatbib]{mnras}

\usepackage{newtxtext,newtxmath}

\usepackage[T1]{fontenc}

\DeclareRobustCommand{\VAN}[3]{#2}
\let\VANthebibliography\thebibliography
\def\thebibliography{\DeclareRobustCommand{\VAN}[3]{##3}\VANthebibliography}

\usepackage{graphicx}
\usepackage{multirow}
\usepackage{graphicx}	
\usepackage{amsmath}	
\usepackage{placeins}
\usepackage{xfrac}
\usepackage{mathtools}

\usepackage{hyperref}
\hypersetup{colorlinks=true, linkcolor=Navy, citecolor=Navy, urlcolor=Navy}

\usepackage{color}

\definecolor{DarkRed}           {RGB}{139,   0,   0}
\definecolor{Red}               {RGB}{255,   0,   0}
\definecolor{Firebrick}         {RGB}{178,  34,  34}
\definecolor{Crimson}           {RGB}{220,  20,  60}
\definecolor{IndianRed}         {RGB}{205,  92,  92}
\definecolor{LightCoral}        {RGB}{240, 128, 128}
\definecolor{Salmon}            {RGB}{250, 128, 114}
\definecolor{DarkSalmon}        {RGB}{233, 150, 122}
\definecolor{LightSalmon}       {RGB}{255, 160, 122}

\definecolor{OrangeRed}         {RGB}{255,  69,   0}
\definecolor{Tomato}            {RGB}{255,  99,  71}
\definecolor{DarkOrange}        {RGB}{255, 140,   0}
\definecolor{Coral}             {RGB}{255, 127,  80}
\definecolor{Orange}            {RGB}{255, 165,   0}

\definecolor{DarkKhaki}         {RGB}{189, 183, 107}
\definecolor{Gold}              {RGB}{255, 215,   0}
\definecolor{Khaki}             {RGB}{240, 230, 140}
\definecolor{PeachPuff}         {RGB}{255, 218, 185}
\definecolor{Yellow}            {RGB}{255, 255,   0}
\definecolor{PaleGoldenrod}     {RGB}{238, 232, 170}
\definecolor{Moccasin}          {RGB}{255, 228, 181}
\definecolor{PapayaWhip}        {RGB}{255, 239, 213}
\definecolor{LightGoldenrodYellow}{RGB}{250, 250, 210}
\definecolor{LemonChiffon}      {RGB}{255, 250, 205}
\definecolor{LightYellow}       {RGB}{255, 255, 224}

\definecolor{Maroon}            {RGB}{128,   0,   0}
\definecolor{Brown}             {RGB}{165,  42,  42}
\definecolor{SaddleBrown}       {RGB}{139,  69,  19}
\definecolor{Sienna}            {RGB}{160,  82,  45}
\definecolor{Chocolate}         {RGB}{210, 105,  30}
\definecolor{DarkGoldenrod}     {RGB}{184, 134,  11}
\definecolor{Peru}              {RGB}{205, 133,  63}
\definecolor{RosyBrown}         {RGB}{188, 143, 143}
\definecolor{Goldenrod}         {RGB}{218, 165,  32}
\definecolor{SandyBrown}        {RGB}{244, 164,  96}
\definecolor{Tan}               {RGB}{210, 180, 140}
\definecolor{Burlywood}         {RGB}{222, 184, 135}
\definecolor{Wheat}             {RGB}{245, 222, 179}
\definecolor{NavajoWhite}       {RGB}{255, 222, 173}
\definecolor{Bisque}            {RGB}{255, 228, 196}
\definecolor{BlanchedAlmond}    {RGB}{255, 235, 205}
\definecolor{Cornsilk}          {RGB}{255, 248, 220}

\definecolor{DarkGreen}         {RGB}{  0, 100,   0}
\definecolor{Green}             {RGB}{  0, 128,   0}
\definecolor{DarkOliveGreen}    {RGB}{ 85, 107,  47}
\definecolor{ForestGreen}       {RGB}{ 34, 139,  34}
\definecolor{SeaGreen}          {RGB}{ 46, 139,  87}
\definecolor{Olive}             {RGB}{128, 128,   0}
\definecolor{OliveDrab}         {RGB}{107, 142,  35}
\definecolor{MediumSeaGreen}    {RGB}{ 60, 179, 113}
\definecolor{LimeGreen}         {RGB}{ 50, 205,  50}
\definecolor{Lime}              {RGB}{  0, 255,   0}
\definecolor{SpringGreen}       {RGB}{  0, 255, 127}
\definecolor{MediumSpringGreen} {RGB}{  0, 250, 154}
\definecolor{DarkSeaGreen}      {RGB}{143, 188, 143}
\definecolor{MediumAquamarine}  {RGB}{102, 205, 170}
\definecolor{YellowGreen}       {RGB}{154, 205,  50}
\definecolor{LawnGreen}         {RGB}{124, 252,   0}
\definecolor{Chartreuse}        {RGB}{127, 255,   0}
\definecolor{LightGreen}        {RGB}{144, 238, 144}
\definecolor{GreenYellow}       {RGB}{173, 255,  47}
\definecolor{PaleGreen}         {RGB}{152, 251, 152}

\definecolor{Teal}              {RGB}{  0, 128, 128}
\definecolor{DarkCyan}          {RGB}{  0, 139, 139}
\definecolor{LightSeaGreen}     {RGB}{ 32, 178, 170}
\definecolor{CadetBlue}         {RGB}{ 95, 158, 160}
\definecolor{DarkTurquoise}     {RGB}{  0, 206, 209}
\definecolor{MediumTurquoise}   {RGB}{ 72, 209, 204}
\definecolor{Turquoise}         {RGB}{ 64, 224, 208}
\definecolor{Aqua}              {RGB}{  0, 255, 255}
\definecolor{Cyan}              {RGB}{  0, 255, 255}
\definecolor{Aquamarine}        {RGB}{127, 255, 212}
\definecolor{PaleTurquoise}     {RGB}{175, 238, 238}
\definecolor{LightCyan}         {RGB}{224, 255, 255}

\definecolor{Navy}              {RGB}{  0,   0, 128}
\definecolor{DarkBlue}          {RGB}{  0,   0, 139}
\definecolor{MediumBlue}        {RGB}{  0,   0, 205}
\definecolor{Blue}              {RGB}{  0,   0, 255}
\definecolor{MidnightBlue}      {RGB}{ 25,  25, 112}
\definecolor{RoyalBlue}         {RGB}{ 65, 105, 225}
\definecolor{SteelBlue}         {RGB}{ 70, 130, 180}
\definecolor{DodgerBlue}        {RGB}{ 30, 144, 255}
\definecolor{DeepSkyBlue}       {RGB}{  0, 191, 255}
\definecolor{CornflowerBlue}    {RGB}{100, 149, 237}
\definecolor{SkyBlue}           {RGB}{135, 206, 235}
\definecolor{LightSkyBlue}      {RGB}{135, 206, 250}
\definecolor{LightSteelBlue}    {RGB}{176, 196, 222}
\definecolor{LightBlue}         {RGB}{173, 216, 230}
\definecolor{PowderBlue}        {RGB}{176, 224, 230}

\definecolor{Indigo}            {RGB}{ 75,   0, 130}
\definecolor{Purple}            {RGB}{128,   0, 128}
\definecolor{DarkMagenta}       {RGB}{139,   0, 139}
\definecolor{DarkViolet}        {RGB}{148,   0, 211}
\definecolor{DarkSlateBlue}     {RGB}{ 72,  61, 139}
\definecolor{BlueViolet}        {RGB}{138,  43, 226}
\definecolor{DarkOrchid}        {RGB}{153,  50, 204}
\definecolor{Fuchsia}           {RGB}{255,   0, 255}
\definecolor{Magenta}           {RGB}{255,   0, 255}
\definecolor{SlateBlue}         {RGB}{106,  90, 205}
\definecolor{MediumSlateBlue}   {RGB}{123, 104, 238}
\definecolor{MediumOrchid}      {RGB}{186,  85, 211}
\definecolor{MediumPurple}      {RGB}{147, 112, 219}

\definecolor{MistyRose}         {RGB}{255, 228, 225}
\definecolor{AntiqueWhite}      {RGB}{250, 235, 215}
\definecolor{Linen}             {RGB}{250, 240, 230}
\definecolor{Beige}             {RGB}{245, 245, 220}
\definecolor{WhiteSmoke}        {RGB}{245, 245, 245}
\definecolor{LavenderBlush}     {RGB}{255, 240, 245}
\definecolor{OldLace}           {RGB}{253, 245, 230}
\definecolor{AliceBlue}         {RGB}{240, 248, 255}
\definecolor{Seashell}          {RGB}{255, 245, 238}
\definecolor{GhostWhite}        {RGB}{248, 248, 255}
\definecolor{Honeydew}          {RGB}{240, 255, 240}
\definecolor{FloralWhite}       {RGB}{255, 250, 240}
\definecolor{Azure}             {RGB}{240, 255, 255}
\definecolor{MintCream}         {RGB}{245, 255, 250}
\definecolor{Snow}              {RGB}{255, 250, 250}
\definecolor{Ivory}             {RGB}{255, 255, 240}
\definecolor{White}             {RGB}{255, 255, 255}

\definecolor{Black}             {RGB}{  0,   0,   0}
\definecolor{DarkSlateGray}     {RGB}{ 47,  79,  79}
\definecolor{DimGray}           {RGB}{105, 105, 105}
\definecolor{SlateGray}         {RGB}{112, 128, 144}
\definecolor{Gray}              {RGB}{128, 128, 128}
\definecolor{LightSlateGray}    {RGB}{119, 136, 153}
\definecolor{DarkGray}          {RGB}{169, 169, 169}
\definecolor{Silver}            {RGB}{192, 192, 192}
\definecolor{LightGray}         {RGB}{211, 211, 211}
\definecolor{Gainsboro}         {RGB}{220, 220, 220}

\definecolor{DavysGray}         {RGB}{ 85,  85,  85}
\definecolor{Jet}               {RGB}{ 52,  52,  52}

\newcommand{\AAA}{\textup{\AA}}

\newcommand{\LSF}{ \mathrm{LSF} }
\newcommand{\II}{ \mathrm{I_2} }
\newcommand{\Te}{ \mathrm{Te_2} }

\newcommand{\kmps}{\textup{km/s}}
\newcommand{\mps}{\textup{m/s}}
\newcommand{\cmps}{\textup{cm/s}}
\newcommand{\ppm}{\textup{ppm}}

\newcommand{\VAL}[1]{{{#1}}}

\newcommand{\Color}[1]{\color{black}}
\newcommand{\Cdot}{}
\newcommand{\mathBF}[1]{#1}
\newcommand{\Boldsymbol}[1]{#1}
\renewcommand{\Cdot}{\cdot}

\title[ESPRESSO Iodine Absorption Cell Experiment ]{Validation of the ESPRESSO Wavelength Calibration Using Iodine Absorption Cell Spectra}

\author[Tobias M. Schmidt et al.]{%
Tobias M. Schmidt,$^{1}$\thanks{E-mail: tobias.schmidt@unige.ch} 
Ansgar Reiners,$^{2}$
Michael T. Murphy,$^{3,4}$
Gaspare Lo Curto,$^{5}$
\newauthor Carlos J. A. P. Martins,$^{6,7}$
Philipp Huke$^{8}$\\
  $^{1}$Observatoire Astronomique de l’Universit\'e de Gen\`eve, Chemin Pegasi 51, Sauverny, CH-1290, Switzerland\\
  $^{2}$Institut für Astrophysik und Geophysik, Georg-August-Universität, 37077 Göttingen, Germany\\
  $^{3}$Centre for Astrophysics and Supercomputing, Swinburne University of Technology, Hawthorn, Victoria 3122, Australia\\
  $^{4}$ARC Centre of Excellence in Optical Microcombs for Breakthrough Science (COMBS)\\
  $^{5}$ESO - European Southern Observatory, Av. Alonso de Cordova 3107, Vitacura, Santiago, Chile\\
  $^{6}$Centro de Astrof\'{\i}sica da Universidade do Porto, Rua das Estrelas, 4150-762 Porto, Portugal\\
  $^{7}$Instituto de Astrof\'{\i}sica e Ci\^encias do Espa\c co, Universidade do Porto, Rua das Estrelas, 4150-762 Porto, Portugal\\
  $^{8}$University of Applied Sciences Emden/Leer, Constantiaplatz 4, Emden Germany\\
}

\date{Accepted XXX. Received YYY; in original form ZZZ}

\pubyear{2025}

\newcommand\blfootnote[1]{%
  \begingroup
  \renewcommand\thefootnote{}\footnote{#1}%
  \addtocounter{footnote}{-1}%
  \endgroup
}

\begin{document}
\label{firstpage}
\pagerange{\pageref{firstpage}--\pageref{lastpage}}
\maketitle

\begin{abstract}
High quality wavelength calibration is crucial for science cases like radial-velocity studies of exoplanets, the search for a possible variation of fundamental constants, and the redshift drift experiment.
However, for state-of-the-art spectrographs it has become difficult to verify the wavelength calibration on sky, because no astrophysical source provides spectra with sufficiently stable or accurate wavelength information.
We therefore propose to use iodine absorption cells to validate the wavelength calibration. 
Observing a bright and featureless star through the iodine cell emulates an astrophysical target with exactly known spectral features that can be analyzed like any other science target, allowing to verify the wavelength calibration derived from the internal calibration sources and to identify systematics in the data processing.
As demonstration, we temporarily installed an $\II$ absorption cell at ESPRESSO.
Employing a full forward modeling approach of the $\II$ spectrum, including the instrumental line-spread function, we demonstrate wavelength calibration \textit{accuracy} at the level of a \VAL{few $\mps$}.
We also show that wavelength measurements do depend on the geometry of the light-injection into the spectrograph fibers. This highlights the importance of probing exactly the same light path as science targets, something not possible with internal calibration sources alone.
We also demonstrate excellent radial-velocity \textit{stability} at the $\VAL{<20\,\cmps}$ level in a full end-to-end fashion, from sky to data product.
Our study therefore showcases the great potential of absorption cells for the verification  and long-term monitoring of the wavelength calibration as well as the unique insights they can provide.
\end{abstract}



\section{Introduction}

\blfootnote{Based on observations collected at the European Southern Observatory under ESO program 60.A-9680(A).}%
Extremely \textit{precise}, \textit{accurate}, and \textit{stable} wavelength calibration of the spectrograph is crucial for several highly ambitious science cases. Most notable here are the hunt for extrasolar planets using the radial-velocity method (RV, \citealt{Griffin1967, Baranne1979, Mayor1983, Mayor1995}), the search for a variation of fundamental physical constants \citep[e.g.][]{Webb1999,Martins2017,Murphy2021}, and the redshift drift experiment, aiming at observing the expansion of the Universe in real time \citep{Sandage1962,Liske2008}. In particular the latter two ones are major science drivers of the ArmazoNes high Dispersion Echelle Spectrograph \citep[ANDES,][]{Marconi2022, Marconi2024, Martins2024}, the future high-resolution spectrograph for the European Extremely Large Telescope \citep[ELT,][]{Padovani2023}. With its huge 39\,m mirror, it will provide an unprecedented light-gathering power and facilitate an approximately 5$\times$ improvement in signal-to-noise (S/N) statistics over current 8 -- 10\,m class telescopes.
It will thus enable obtaining substantially tighter constraints on the fine-structure constant and for the first time provide sufficient photons to realistically attempt the redshift-drift experiment.
However, these science cases crucially rely on the quality of the spectrograph's wavelength calibration, because the underlying measurement principle
revolves around a highly precise analysis of spectral absorption features.
Given that systematics are already a critical issue for current observations, it is mandatory to also achieve a similar improvement in terms of instrumental systematics and in particular wavelength calibration, to actually benefit from the larger collecting area of the ELT and the hugely improved photon statistics.

Observations of quasar spectra currently constrain the fine-structure constant, $\alpha=e^2/\hbar{}c$, at the level of approximately one part-per-million, i.e. $\Delta{}\alpha/\alpha \approx 10^{-6} = 1\,\ppm$. This is achieved by a
differential wavelength measurement of multiple absorption lines in the same spectrum and
therefore requires the wavelength calibration to be \textit{accurate}%
\footnote{Throughout this paper, the terms \textit{precision}, \textit{accuracy}, and \textit{stability} are used as defined in \citet{Martins2024} and described more extensively in the appendix of \citet{Schmidt2024}. }
and in particular free of distortions.
A variation of $\alpha$ by 1~ppm causes differential line shifts of less than $25\,\mps$, a small fraction of a spectral resolution element, which typically encompasses about $3\,\kmps$ and highlights the challenge associated with this type of measurements.
Several studies \citep[e.g.][]{Griest2010, Whitmore2010, Whitmore2015} have demonstrated that wavelength solutions of slit spectrographs can easily be affected by distortions of several hundred $\mps$, raising substantial doubts about the early results derived from observations with Keck/HIRES and VLT/UVES \citep[e.g][]{Webb1999, Murphy2003, Quast2004,  Chand2006, King2012, Molaro2013a, Kotus2017}. 
Massive improvements in terms of systematics have recently been achieved using fiber-fed spectrographs \citep{Milakovic2021, Murphy2021}, but even for these instruments wavelength calibration systematics can become an issue when aiming for constraints at the 1~ppm level and better \citep[e.g.][]{Schmidt2021}. To ensure that the 5$\times$ more precise observations obtained with ANDES will be photon-limited and not significantly affected by instrument systematics, one aims for a wavelength calibration \textit{accuracy} at the $1\,\mps$ level \citep{Martins2024}, about one order of magnitude more demanding than achievable with current instruments.

The redshift drift, on the other hand, is in principle a classical RV experiment, very similar to the search for exoplanets, and requires \textit{stability} of the wavelength calibration. Following the standard $\Lambda$CDM model of cosmology \citep{Planck2018}, the expected signal caused by cosmic expansion will\,--\,depending on the redshift\,--\,amount to just a few $\cmps$ per decade, far less than the current state-of-the-art in RV exoplanet observations (Proxima~Cen~d, $39\,\cmps$, \citealt{Faria2022}). The linear drift will also be much harder to detect and easier to be confused with instrumental effects than the periodic RV signal induced by exoplanets.
The goal for ANDES is therefore to ensure a wavelength calibration \textit{stability} of $1\,\cmps$ over the lifetime of the instrument \citep{Martins2024}, a tremendous challenge and far in excess of everything demonstrated so far.

Achieving these extremely ambitious goals will require a community effort over the next decade and advancements in many different fields related to wavelength calibration.
Essential will be to have a powerful calibration unit, equipped with several sophisticated calibration sources. These will include classical thorium-argon and uranium-neon hollow-cathode lamps \citep[ThAr, UNe HCL, e.g.][]{Kerber2007, Nave2018, Sarmiento2018}, combined with white-light Fabry-P\'erot etalons \cite[FP,][]{Perot1899}. These are by now well-established as high-fidelity calibrators for echelle spectrographs \citep[e.g.][]{Wildi2010,Wildi2011,Wildi2012, Bauer2015, Stürmer2017, Seifahrt2018, Hobson2021}.
A fully complementary technology with great potential for calibration of astronomical spectrographs are laser frequency combs \citep[LFC, e.g.][]{Steinmetz2008, Wilken2010a, Wilken2012, Phillips2012a, Probst2014,Probst2016, Kokubo2016, Metcalf2019a}.
These produce a plethora of narrow lines with wavelengths linked directly to the fundamental SI frequency standard.
However, due to their inherent complexity, frequent reliability issues, and the difficulty to make full use of their potential, these devices have so far seen only limited use in practice for regular calibration of astronomical observations \citep[e.g.][]{Hirano2020, Blackman2020, Terrien2021}. There is therefore a strong need to bring this technology to the maturity level required for daily, unsupervised operations at an observatory.
In addition, efforts are ongoing to develop novel technology extending the spectral range of LFCs and allow to use them for spectrograph calibration also in the blue and near-UV range \citep[e.g.][]{Cheng2024, Wu2024, Ludwig2024}.
Apart from the calibration sources, software and algorithms used to process and calibrate the data will be of crucial importance.  Fairly capable methods do exist
\citep[e.g.][]{Zechmeister2014, Piskunov2021, Zhao2021, Cook2022, Schmidt2024}, but thorough investigations and dedicated test have nevertheless revealed a large variety of different kinds of systematics that are ultimately related to an imperfect extraction and calibration process which is not capable of entirely capturing all instrumental effects present in the raw data \citep[see e.g.][]{Bouchy2009,Bolton2010,Zechmeister2014,Probst2016,Blackman2020,Milakovic2020a, ZhaoF2021, Schmidt2021,Schmidt2024}.
Reaching the necessary improvement over the current state of the art might therefore require to develop or implement even more advanced algorithms \citep[e.g.][]{Bolton2010, Guy2023}.
Achieving the ANDES wavelength calibration goals for accuracy and stability will in any case pose a tremendous challenge but there are promising paths and ongoing efforts to push the current limits and make notable improvements.

\begin{figure}
 \includegraphics[width=\linewidth]{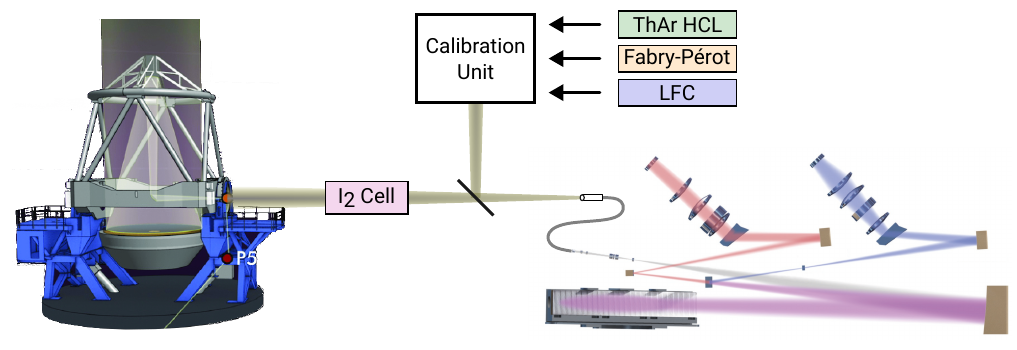}
 \caption{  Illustration of the simplified light path of ESPRESSO \citep{Pepe2021}, indicating the placement of the different calibration sources. Light from the ThAr HCL, the FP, or the LFC is injected at the instrument front-end into the spectrograph fibers via a retractable mirror. The absorption cell, however, is placed in the beam coming from the VLT unit telescope and therefore imprint the I$_2$ lines onto the spectrum of the observed target, in this case a nearly featureless and rapidly rotating star.}
 \label{Fig:Sketch1}
\end{figure}

However, there arises the question of how to properly \textit{validate} the achieved wavelength calibration, in particular at the very demanding level of accuracy and stability required for the ANDES key science cases.
Such a validation must necessarily also include all data processing steps, i.e. the extraction of science and calibration spectra, the establishment of wavelength solution, the modeling of the instrumental line-spread function (LSF), and the analysis of the data itself. These efforts, however, will be essential to build confidence in the wavelength calibration process and give credibility to the scientific results by demonstrating that observed effects are indeed of astrophysical origin and not caused by spurious instrumental effects.
To do so, one can of course compare different calibration sources against each other, for example lines produced by HCLs, LFCs, or FPs, but also the  derived ThAr/FP and LFC wavelength solutions \citep[see e.g][]{Probst2016, Milakovic2020a, Schmidt2021, Terrien2021, Schmidt2024, Ludwig2024}.
One can also check for the consistency of such a comparisons in different spectral orders (in regions where they overlap), or in spectra obtained from different spectrograph fibers or slices to identify internal inconsistencies \citep{Schmidt2021, Schmidt2024}. However, even in absence of such discrepancies, these kind of tests can not provide a true end-to-end validation of the wavelength calibration process.
A rigorous test of \textit{accuracy} is in general hardly possible because a ground-truth wavelength standard is not available. A more practical issue, however, is that light provided by the calibration unit does not follow the exact same path as science observations. It is typically injected into the spectrograph fiber at the front-end and does not pass through the telescope (see Figure~\ref{Fig:Sketch1}).
To be fully representative, the optical path for a verification of the wavelength calibration should be identical to science observations, i.e. the source should be located on sky and the light pass through atmosphere and telescope before reaching the spectrograph. A particular issue is that differences in the fiber-injection geometry can propagate and have a direct impact on the measured wavelengths \citep{Hunter1992, Chazelas2012}.
In addition, to avoid circularity of the argumentation and provide verification for all sorts of observed targets, the validation source should provide spectral features that are dissimilar from the ones in the calibration sources and\,--\,if possible\,--\,similar to the features in the science spectra. Most calibration sources, however, provide emission-style lines (e.g. ThAr, FP, or LFC sources) while the features analyzed in science observations are mostly absorption lines. At the level required for ANDES, this can make a difference, in particular when certain types of systematics are present in the data reduction and calibration process.

Historically, one has validated the stability of spectrographs on-sky by observing stable RV reference stars \citep[e.g.][]{Pepe2021}.
However, there exists no astrophysical source that matches the increased ambitions of current or future extreme-precision RV spectrographs.
The probably most stable RV reference star, $\tau$~Ceti, exhibits an intrinsic RV noise between $70\,\cmps$ and $1\,\mps$, 
in excess of the $1\,\cmps$ ANDES stability goal by nearly two orders of magnitude.
Using the Sun as a reference, although providing ample photons, is no solution either, because its inherent RV noise is at the $\mps$-level \citep[e.g.][]{Dumusque2021,Zhao2023}.
Careful analysis of absorption lines imprinted onto the spectra by Earth's atmosphere can in principle be used to assess the spectrograph stability \citep[e.g.][]{Griffin1973, Figueira2010a, Figueira2010b}. However, this method is only applicable in the near-IR where telluric lines are abundant and limited to few $\mps$.
Also exotic means of wavelength calibration have been suggested, e.g. observing the scattered light from laser-guide stars \citep{Vogt2018,Vogt2019,Vogt2023} or placing artificial calibration sources on drones or satellites, but were so far not adopted or did not materialize at all.

Here, we instead propose and demonstrate in practice a simple, reliable, and effective strategy for a representative end-to-end \textit{validation} of the wavelength calibration of high-resolution echelle spectrographs using iodine absorption cells.
The concept is sketched in Figure~\ref{Fig:Sketch1}.
The goal is clearly not to employ $\II$ absorption cells to \textit{calibrate} science observations \citep[as pioneered e.g. in][]{Marcy1992, Butler1996}. This would be rather inefficient for various reasons, e.g due to the massive cluttering of the science spectrum with contaminating lines imprinted by the cell, significant absorption of valuable science photons, and the excessive blending of iodine lines.
Wavelength calibration of the spectrograph is certainly done in a better and more efficient way using engineered calibration sources, e.g. following the proven strategies for establishing ThAr/FP or LFC based wavelength solutions \citep[see][]{Bauer2015, Cersullo2017, Cersullo2019, Schmidt2021, Hobson2021, Schmidt2024}.
Instead, observations through absorption cells shall exclusively be used to \textit{validate} the wavelength calibration of the spectrograph.
This is done by observing bright, featureless, and fast rotating stars through an iodine absorption cell.
In this setup, the star just acts as a light source, provides the photons, and ensures that the same optical path is probed as for actual science observations.
All spectral information, however, are imprinted by the absorption cell, which provides highly stable and accurately known spectral features \citep[e.g.][]{Knöckel2004,Debus2023, Reiners2024}.
In addition, the absorption cell makes the $\II$ spectrum accessible to be measured in the future with even higher precision if needed, using e.g. Fourier-transform spectrometers (FTS) or various flavors of laser spectroscopy, which can in principle provide absolute frequency accuracy.

In the described setup, observations through the absorption cell basically emulate a science target, however, with exactly known spectral features.
Following the observations, the $\II$ spectra can be processed by the standard data reduction and calibration pipeline, analyzed similarly to science observations, and the inferred wavelengths compared to their true values.
Here, iodine cells produce absorption lines over the wavelength range from approx. 5000 to $6300\,\mathrm{\AAA}$,  which are more similar to the typical features of interest in science observations than the emission lines provided by most calibration sources.
In addition, the $\II$ absorption lines are narrow, unresolved, and heavily blended, which in terms of instrument systematics (e.g. w.r.t. the LSF) poses a particularly strong challenge to the wavelength calibration and analysis procedure.
Apart from the variable barycentric correction, which cannot be emulated in this way, the
proposed scheme enables a true end-to-end validation, from sky to final data product, including the wavelength calibration of the spectrograph but also the full data processing procedures. By this, it is very closely matched and highly representative for actual science observations.

The technological and operational challenges related to $\II$ cells are minimal, provided that some part of the telescope beam can be accessed and the cell placed there.
The iodine spectrum is by now thoroughly characterized and accurately described by models of the atomic energy levels and associated transitions \citep[e.g.][]{Knöckel2004, Perdelwitz2018, Debus2023, Reiners2024}.
In addition, substantial advancements have recently been achieved concerning an accurate characterization of instrumental line-spread functions \citep[e.g.][]{Hirano2020,Schmidt2024}, which is essential for a proper forward-modeling of the observed iodine absorption spectrum.
All necessary ingredient to employ the described concept for validation of the wavelength calibration with $\II$ cells are therefore available.
We thus have temporarily installed an iodine absorption cell at the VLT/ESPRESSO spectrograph in May 2023 and performed numerous tests to demonstrate the feasibility of the outlined validation scheme in practice and test the accuracy and stability of the ESPRESSO wavelength solution.

\section{Instrumental Setup and Observations}
\label{Sec:Data}

ESPRESSO is the Echelle SPectrograph for Rocky Exoplanets and Stable Spectroscopic Observations (\citealt{Megevand2014, GonzalezHernandez2018, Pepe2021}).
It is installed at the ESO Very Large Telescope (VLT) since 2018 and the current flagship for astronomical observations that require particularly accurate or stable wavelength calibration.
Its primary science drivers are the hunt for Earth-sized extrasolar planets and the search for a possible variation of fundamental physical constants, hence the special emphasis in the design on a high-quality wavelength calibration.
ESPRESSO offers a spectral resolving power of $R = \frac{\lambda}{\Delta{}\lambda}\approx 135\,000$ in the standard \texttt{HR} observing mode and incorporates numerous techniques to achieve highly stable and accurate wavelength calibration. For instance, the full instrument is enclosed in a vacuum vessel, thermally stabilized to the milli-Kelvin level, fiber-fed, and equipped with a full suite of calibration sources, encompassing a ThAr HCL, a highly-stable white-light FP etalon \citep[e.g.][]{Wildi2011}, and a LFC, similar to the one described in \citet{Probst2014,Probst2016}.
To achieve the desired high resolving power and limit the physical size of the echelle grating, ESPRESSO employs a pupil slicer in its design \citep{dellAgostino2014}. Therefore, the spectra from each of the two input fibers are imaged twice onto the detector. These can be extracted separately and thus offer the possibility for cross-checks between the formally identical spectra obtained from both slices.

A unique feature of ESPRESSO is that it can be fed by any or all four VLT Unit Telescopes \citep{GraciaTemich2018b, Pepe2021}. For this, light from the telescopes is channeled through the Coud\'e trains to a joint Incoherently Combined Coud\'e Focus located in the underground ICCF laboratory \citep{Cabral2010, Cabral2012, Cabral2014} where it is injected into the spectrograph fibers by the instrument front-end \citep{Riva2014b}. After exiting the Coud\'e tunnels, the beams propagate freely through the ICCF laboratory for several meters before entering the ESPRESSO front-end structure.
This location is therefore ideal to conveniently place an absorption cell into the optical path.
The beams coming from the telescopes are in this region slowly converging with an f-number of $f/22.8$ and have a diameter of about 36\,mm before entering the front-end structure.
Therefore, 3" cells are sufficient.


At Paranal, three cells are available.
One cell is empty and was used to represent the additional optical components in the beam during reference observations taken without iodine absorption. However, no significant change of the focus position of the Coud\'e train could be detected. The optical effects caused by the 
cell windows must therefore have been minimal.
Two other cells are filled with iodine at fill temperatures of \VAL{36} and \VAL{37$^\circ\textup{C}$}. Here, the $\II$ vapor pressure at the fill temperature defines the amount of iodine contained in the cells and therefore the strength of the absorption.
During operations, the cells are heated to substantially higher temperatures, e.g. between 60$^\circ\textup{C}$ and 70$^\circ\textup{C}$, which ensures that all iodine is in the gaseous phase, without any condensations remaining.
Critical here is often the temperature at the filling neck of the cell and at the entrance windows, which are naturally colder and have a tendency to accumulate condensed iodine.
The experiment was conducted with the cell filled at \VAL{36$^\circ\textup{C}$}, wrapped in heating foil and some insulation foam.
Two thermocouples on the cell, connected to a  simple bench-top temperature controller, facilitated temperature control.
The temperature regulation of the cell was not integrated into the instrument control software and no dedicated logs of the cell temperature are available.
Therefore, the actual temperature of the cell is not exactly known, but accurate control of the cell temperature during operations has anyway been a common issue in the past, see e.g. \citet{Perdelwitz2018} or \citet{Wang2020}. However, this aspect is also not considered particularly critical.
As long as all parts of the cell stay safely above the fill temperature, no condensation occurs and the strength of the observed $\II$ absorption should therefore be rather insensitive to the actual temperature.
When in use, the cell was manually placed on a simple fixture in the optical beam coming from UT2. 
This was sufficient, since no special alignment of the absorption cell or an exact reproducibility is necessary.
The manual movement of the cell, however, required regular access to the ICCF laboratory, e.g. for switching between spectra taken through empty and filled cell or before handing over the instrument for regular nighttime observations.


The iodine absorption cell experiment ran for two weeks from May 11 to May 25, 2023.
The first days were used for setting up the equipment and defining the operational procedures for these non-standard observations. Some issues with the temperature control of the cell had to be solved during this period.
There was a tendency for $\II$ condensation on one of the cell windows. This leads to reduced $\II$ absorption, which could later also be confirmed from analyzing the spectra.
The setup was fully operational from the 16th onward.
Some cloud coverage in the following days, however, prohibited observations, leaving just six days for regular operations.

All on-sky observations were conducted during the brief twilight periods, mostly in the evenings. Here, $\II$ observations had to be completed and the cell removed from the optical path before regular nightime operations with ESPRESSO commenced.
Some data were taken during morning twilights, but this proved to be inconvenient for logistical and operational reasons.
On-sky twilight observations were complemented by some longer sequences taken during the day and using a lamp in the ICCF laboratory as light source.

During the course of the experiment, the use of several instrument modes was explored and different stars targeted.
However, we focus in the following on observations taken in \texttt{1HR2x1} mode, i.e. with a resolving power of $R\approx135\,000$ and 2$\times$1 binning (spatial $\times$ spectral direction), with simultaneous FP calibrations on Fiber~B (instead of sky), and targeting the fast-rotating and nearly featureless star HD\,69081.
It is of spectral type B2 and has a brightness of $V \approx 5\,\mathrm{mag}$, sufficient to reach a good signal-to-noise ratio (S/N), i.e. about half the full-well capacity of the detector, in a bit over 100\,seconds. This target is therefore well-suited for our endeavor and was predominantly observed. We stress, however, that many other equally good stars exist as well and  that there are plenty of targets to choose from.
The selected subset of the available data is in terms of instrument mode fully representative for actual quasar observations and most informative in terms of wavelength calibration.

\section{Data Analysis}
\label{Sec:Analysis}

\begin{figure}
 \includegraphics[width=\linewidth]{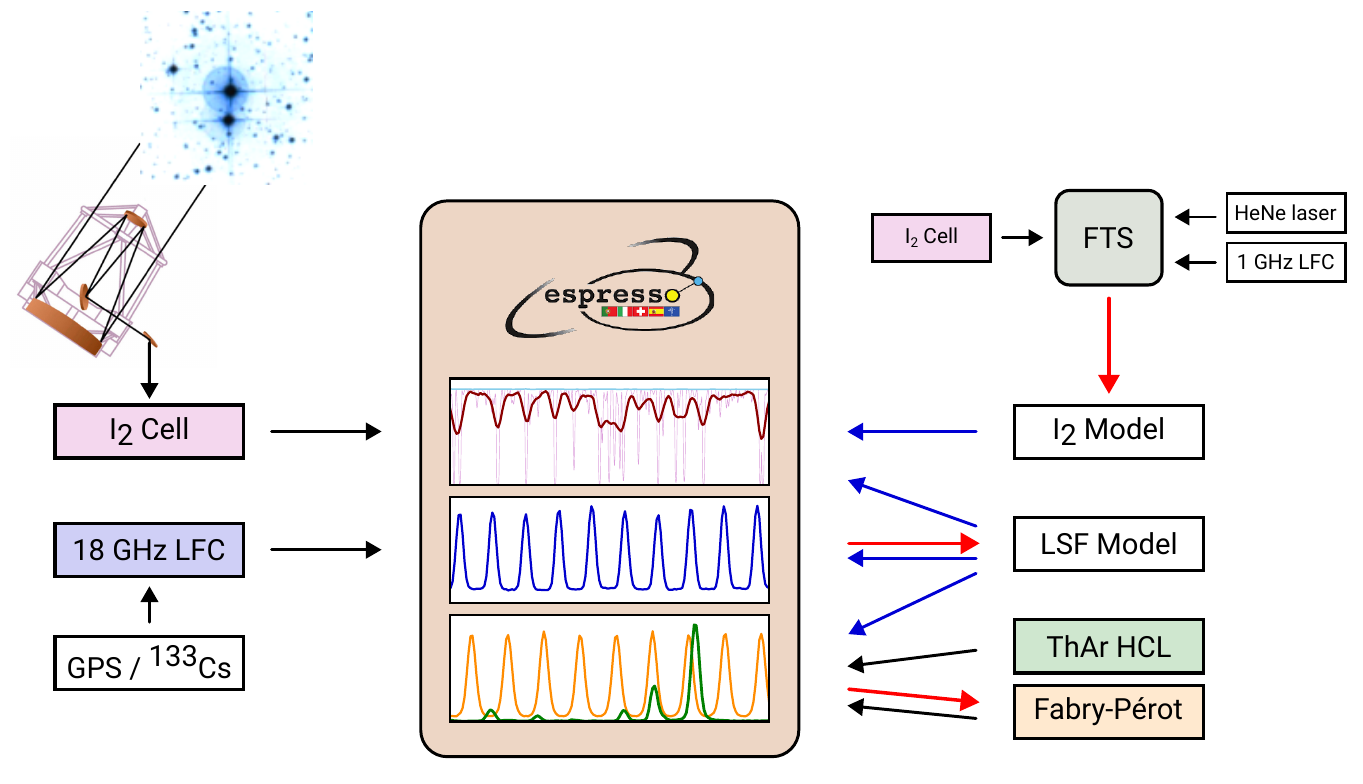}%
 \caption{
    Schematic representation of the different types of spectra recorded by ESPRESSO and their mutual relation within the context of this analysis. A detailed description is given in Section~\ref{Sec:Analysis}.
    Black arrows indicate light paths, red ones where observed data is used to inform a model, and blue arrows where model information contributes to the processing of the observations.
    The spectra displayed in the center (HD\,69081 through $\II$ cell, LFC, and FP\,+\,ThAr, from top to bottom) are actual observations plotted over the identical wavelength interval around $5255\,\mathrm{\AAA}$. The top panel also shows the corresponding best-fitting $\II$ model prior to convolution with the LSF (thin line). The FTS-based super-calibration of the $\II$ model is indicated in the top-right corner.}
  \label{Fig:Sketch2}
\end{figure}

Validation of the ESPRESSO wavelength calibration is performed by comparing the standard ThAr/FP or LFC wavelength solution to the wavelength scale of a semi-empirical $\II$ model that is used to fit the iodine absorption spectra. The overall calibration and validation process is rather complex and the general concept visualized in Figure~\ref{Fig:Sketch2}.
Key aspect of the validation is that spectra of sources with rather different properties can be observed with the same, highly stable spectrograph.
A fundamental requirement for any spectrograph that makes use of dedicated calibration exposures is of course that various kinds of science targets as well as the calibration sources are imaged identically onto the detectors and are recorded in an equivalent and comparable way. This requirement drives the design of the fiber feed, which is supposed to scramble the input light, provide a stable and homogeneous illumination of the spectrograph, ensure comparability between measurement of different sources, and actually makes the instrument calibratable. The absorption cell experiment carried out here allows to test and verify this requirement.
The variety of sources shown in Figure~\ref{Fig:Sketch2} offers the opportunity for numerous cross-comparisons and allows to benefit from the particular properties of each of these sources.
For instance, the LFC provides a huge number of narrow and truly unresolved lines which are used to construct a non-parametric model of the instrumental LSF. This is then subsequently used to model spectral features in the analysis of all types of spectra.
The classical ThAr/FP wavelength calibration is constructed from the thorium-argon HCL spectra, which provide the absolute wavelength information, and the FP, which delivers regularly spaced lines and therefore allows accurate interpolation over the full wavelength range of the spectrograph.
A fully independent wavelength solution is derived from the LFC. Here, the absolute wavelength information comes from the lock of offset frequency and repetition rate to the international SI time standard via a GPS GNSS receiver.
Iodine absorption spectra, obtained by observations of a bright, featureless star through the $\II$~cell are modeled using a semi-empirical model of the iodine ro-vibronic potential levels and therefore connected to an independent wavelength scale. This $\II$ model was calibrated in the laboratory using a FTS that is itself tracked and calibrated by a helium-neon laser and a dedicated LFC.
A more detailed description of the individual aspects is given below.

\subsection{Data Reduction}

All data processing, starting directly from the raw frames, is done by means of a custom data-reduction pipeline \citep{Schmidt2021,Schmidt2024}.
Spectral extraction is performed using a variant of the \textit{flat-relative optimal extraction} algorithm presented by \citet{Zechmeister2014} and therefore naturally flatfield- and blaze-corrected.
Special care was given to accurate bias and gain calibration, as well as treatment of scattered light, although these aspects are probably not particularly critical for the high-S/N spectra used within the context of this study.
Spectra of on-sky targets are flux calibrated using observations of spectrophotometric standard stars.
In the following, extracted spectra of individual spectral orders, separate for both fibers and slices, are used. No merging of spectral orders is performed to preserve the full information content, not mix spectra with different LSFs, and avoid correlating the noise.

\subsection{Wavelength Calibration of the spectrograph}
\label{Sec:SpectrographCalibration}

The wavelength calibration of the spectrograph, including the determination of the LSF, is done in a fully identical way as described by \citet{Schmidt2024}.
Spectra of the LFC, which contain a plethora of narrow, truly unresolved lines and therefore reveal the instrumental profile of the spectrograph, are used to infer a non-parametric model of the ESPRESSO LSF. This is done separately for each spectrograph fiber and slice, and in 16 blocks per echelle order. The suite of LSF models is then used to fit the lines in the ThAr, FP, and LFC spectra.
Following this, the thorium lines are used to calibrate the effective gap size of the FP etalon, $D_\mathrm{eff}(\lambda)$, again in a non-parametric fashion (see \citealt{Schmidt2021} for details). This calibration is repeated for every set of daily wavelength calibration spectra and therefore automatically takes care of the chromatic FP drift \citep{Terrien2021, Schmidt2022, Kreider2022, Basant2025}.
The lines of the calibrated FP are then utilized to derive the combined ThAr/FP wavelength solution for each spectral order, i.e. a relation $\lambda(y)$ between pixel positions along the main dispersion direction, $y$, and the wavelength \citep{Schmidt2024}. We stress here that this wavelength solution can not be taken solely by its own, but always has to be used in conjunction with the corresponding LSF model that was assumed when constructing the wavelength solution.
In a similar, but fully independent way, the LFC lines are used to construct the LFC wavelength solution, relying on the values for offset frequency and repetition rate provided by the LFC system and reported in the \texttt{fits}~headers to calculate the absolute frequency of each LFC line.

\subsection{Model of molecular iodine absorption}
\label{Sec:I2Model}

Modeling of the iodine absorption spectrum is based on an analytic description of the ro-vibronic structure of the $\II$ B-X spectrum, derived from the molecular potentials of the two electronic states as well as their hyperfine parameters, and informed by high precision measurements of the $\II$ B-X spectrum in the visible \citep{Knöckel2004, Reiners2024}.
The model consists of a large list of $\II$ absorption lines. These were included regardless of their predicted absorption strength, which leads to more than \VAL{4\:000\:000} lines in the range $\VAL{5150}$ -- $\VAL{6300}\,\mathrm{\AAA}$, i.e. on average between \VAL{30} and \VAL{100} lines per $1\,\mathrm{\kmps}$ spectral width or \VAL{65} -- \VAL{230} lines per ESPRESSO resolution element ($R \approx 135\,000$).
The expected accuracy of the transition frequencies is better than $\mathrm{\VAL{3\,MHz}}$ in the wavelength range $\VAL{5260}$ -- $\VAL{6670}\,\mathrm{\AAA}$, corresponding to $\mathrm{\VAL{\approx2\,\mps}}$.
Relative line intensities are predicted based on the assumed temperature of the gas, in this case $T = \VAL{60}^\circ\textup{C}$.
We scale all line intensities from the model calculations by a factor of \VAL{3800} to approximately match the absorption strength in the utilized absorption cell.
From the line list, we construct a model spectrum by superposition of thermally Doppler broadened lines, according to the assumed temperature.

For this model of the $\II$ absorption spectrum, \citet{Reiners2024} have presented an additional calibration of the wavelength scale based on a comparison of the model to accurate FTS observations of an iodine cell (see their Figure~4). This correction reaches up to approximately $\VAL{\pm3\,\mps}$ within the wavelength range from $\VAL{5200}$ to $\VAL{6300}\,\mathrm{\AAA}$ and is facilitated by a simultaneous calibration of the FTS by a LFC. For our analysis, we make use of this additional calibration and confirm that it indeed improves the agreement between $\II$ wavelength scale and ESPRESSO wavelength solution.
We stress here that the iodine cell as well as the LFC system used by \citet{Reiners2024} to super-calibrate the $\II$ model in the laboratory are completely different from the equipment used at Paranal. The correction is therefore not specific to the cell used with ESPRESSO but indeed applies to the model itself and should then be universally applicable to any $\II$ absorption cell.

Historically, templates derived from FTS scans of the actual cells have been used to model the $\II$ absorption \citep[e.g.][]{Marcy1992,Butler1996,Wang2020}.
However, the analytical description of the $\II$ spectrum used here has numerous advantages:      
The model represents the iodine spectrum at infinite spectral resolution, is noise-free, and expressed in terms of transmission, i.e. with a perfectly known continuum. For FTS scans, a proper continuum fit would first have to be performed.
Also, although providing very high resolution and excellent linearity of the wavelength scale, FTS have difficulties providing an absolute wavelength scale.
The offset in the optical paths between science spectrum and tracking laser leads to a stretch of the wavelength scale or equivalently a shift in velocity of the whole spectrum which can easily encompass several hundred $\mps$.
This is typically corrected by close inspection of features in the to-be-measured spectrum for which accurate frequency information are available from external sources.
The measurements by \citet{Reiners2024} circumvent this issue by simultaneous calibration of the FTS with a 1~GHz Titanium-sapphire LFC.
Due to the inherent nature of the Fourier transform principle, all wavelengths are always observed at the same time. Thus, calibrations derived from a highly accurate LFC sent through the FTS simultaneously and along the identical light path but in a different spectral domain do apply to the full wavelength range and therefore allow to accurately calibrate the $\II$ spectrum.
Relying on the calibrated iodine model from \citet{Reiners2024}, the ESPRESSO wavelength solution could thus be tested in absolute terms.

\subsection{Fit of the Iodine Absorption Spectra}
\label{Sec:FitProcedure}

To connect the wavelength scale provided by the $\II$ model with the observed iodine absorption spectrum, we perform a fit of the model to the data.
This fit is executed independently in many small chunks and also separately for the two spectrograph slices.
As for the determination of the LSF, we chose 16 blocks per echelle order and assume that the instrumental LSF is constant over each block.
For each block, the fit can be described by the following relation:
  \begin{equation}
   {\Color{MediumBlue}F_\mathrm{cell}}(\,\lambda^\mathrm{spec}\,) \; \xLeftrightarrow{\hspace{.3em}\mathsf{fit}\hspace{.3em}} \; \mathBF{f_0} \; \cdot \; {\Color{DarkOrchid}F_\mathrm{clear}}(\,\lambda^\mathrm{spec}\,) \; \Cdot \;  \left(\;  {\Color{DarkRed}\II} \ast {\Color{DarkOrange}\LSF} \; \right)(\,\lambda^\mathrm{spec}\,) \; .
   \label{Eq:Model1}
  \end{equation}
Here, ${\Color{MediumBlue}F_\mathrm{cell}}(\,\lambda^\mathrm{spec}\,)$ denotes the observed absorption spectrum and $\lambda^{\mathrm{spec}}$ indicates that the spectrum is expressed in terms of the spectrograph's observed-frame wavelength calibration, i.e. either the ThAr/FP or LFC wavelength solution. This includes a component representing the intra-day  drift.
The actual wavelength solutions are established once per day from the wavelength calibration frames taken as part of the regular instrument calibration plan. Over the day, the spectrograph might drift. This relative drift between the calibration exposures (taken in the morning) and the $\II$ absorption exposures (typically during evening twilight) is tracked by the simultaneous FP spectrum, always recorded in parallel by the second spectrograph fiber.
To determine the intra-day drift, we make use of the very efficient flux gradient method \citep{Bouchy2001} and typically apply one average drift correction per spectral order, fiber, and slice. The FP itself is sufficiently stable (typical drift of a few $\cmps$ per day, \citealt{Schmidt2022}) that its own intra-day drift can be ignored.

To describe the observed spectra, a composite model is constructed and fitted to the data.
First ingredient of the model is a high-S/N co-addition of multiple spectra of the target star taken through the empty cell, denoted in Equation~\ref{Eq:Model1} as ${\Color{DarkOrchid}F_\mathrm{clear}}(\,\lambda^\mathrm{spec}\,)$. This provides the overall spectral flux distribution of the source star and contains the (very few) absorption lines originating in the stellar atmosphere.
To account for the variable barycentric correction, the combination of multiple exposures and different spectral orders has to be performed in the barycentric frame and the combined spectrum shifted back to the observed-frame wavelength scale appropriate for the date and time of the $\II$ observation. This, however, is not particularly critical since the stellar spectrum is nearly featureless and not affected by any relevant telluric absorption lines within the spectral range of interest.

To account for intensity differences between the spectra taken through the empty cell and the actual $\II$ absorption observation, a re-scaling factor, $\mathBF{f_0}$, is introduced and optimized during the fitting process. A possible origin for such a difference is the variable seeing which leads to different fiber-injection losses. These are accounted for during the flux calibration process using the scheme presented in \citet{Schmidt2024b}, but the correction is\,--\,as expected\,--\,imperfect and some additional scaling is necessary.

The other part in Equation~\ref{Eq:Model1} deals with the absorption features imprinted by the iodine cell. Here, $\left(\;  {\Color{DarkRed}\II} \ast {\Color{DarkOrange}\LSF} \; \right)(\,\lambda^\mathrm{spec}\,)$ denotes the convolution of the theoretical $\II$ absorption model described in Section~\ref{Sec:I2Model} with the LSF of the spectrograph as determined in Section~\ref{Sec:SpectrographCalibration}.
The formally correct way would be to first construct a de-convolved model of ${\Color{DarkOrchid}F_\mathrm{clear}}$, multiply it with the $\II$ absorption model, and then convolve both with the LSF. However, the stellar spectrum is to fairly good approximation a flat continuum and can therefore be kept separately, outside the convolution operation, which greatly simplifies Equation~\ref{Eq:Model1}. The very few regions with stellar absorption lines where this approximation might be inaccurate can easily be excluded from the analysis, if necessary.

The iodine model is expressed as function of wavelength, while the LSF is defined in velocity space. Relying on the Doppler-formula, the convolution operation can therefore be computed as
\begin{equation}
  \left(\;  {\Color{DarkRed}\II} \ast {\Color{DarkOrange}\LSF} \; \right)\left(\,\lambda^\mathrm{spec}_i\,\right) =
  \int_{-\infty}^{\infty} \; {\Color{DarkRed} \II\, \left({ \color{black} \, \lambda^\mathrm{mod} \Cdot \left(1 - \frac{v}{\mathrm{c}}\right) \, }\right)^{\:\color{black}\Boldsymbol{\tau}}} \hspace{.5em} {\Color{DarkOrange} \LSF(\, v \,) } \; dv .
  \label{Eq:Model2}
\end{equation}
Here, the output of this operation needs to be expressed as function of $\lambda^{\mathrm{spec}}$. The theoretical iodine model, ${\Color{DarkRed} \II\, \left({ \color{black} \, \lambda^\mathrm{mod} }\right)}$, however, comes with its own calibrated wavelength scale, $\lambda_\mathrm{mod}$.
These two wavelength scales do not need to be identical but might relate by
\begin{equation}
 \lambda^{\mathrm{spec}} = \lambda^{\mathrm{mod}}  \Cdot \left( 1 + \frac{\mathBF{\color{black}\Delta{}v}}{\mathrm{c}} \right) \; .
 \label{Eq:RelationWavelengthScales}
\end{equation}
Here, the velocity shift, $\mathBF{\Delta{}v}$, between the two wavelength scales is the key parameter determined during the fitting process and the primary quantity of interest within the context of this analysis. It provides a direct measure of the consistency between the two wavelength scales and its time-evolution gives information about the stability of the system.
The sign convention is chosen in the way that a positive $\mathBF{\Delta{}v}$ indicates that the iodine model has to be redshifted to match the wavelengths observed in the ESPRESSO spectrum. Alternatively, if one assumes that the $\II$ calibration is correct, positive values of $\mathBF{\Delta{}v}$ can be interpreted as that the wavelengths of spectral features determined by ESPRESSO are in fact too large, i.e. too red.

The final model parameter is an adjustment of the $\II$ absorption strength.
The theoretical model is computed for an arbitrary absorption strength and therefore has to be adjusted to the observations.
This is done with the parameter $\Boldsymbol{\tau}$ in Equation~\ref{Eq:Model2}, which effectively re-scales the optical depth of the $\II$ absorption.

Although expressed as integral in Equation~\ref{Eq:Model2}, the convolution operation is in practice computed as a discrete sum over velocity bins, given that the empirical model of the LSF is defined in a non-parametric fashion over a discrete velocity grid \citep{Schmidt2024}.
At the typical temperatures of the cell, the $\II$ molecules in the gas should have a velocity dispersion and therefore cause a Doppler broadening of at least $100\,\mps$. Given the sampling of the LSF in bins of $\VAL{10\,\mps}$ width, it is justified to just evaluate Equation~\ref{Eq:Model2} at the various values of $v$ instead of actually performing the integral over the individual velocity bins.

\begin{figure*}
 \includegraphics[width=\linewidth]{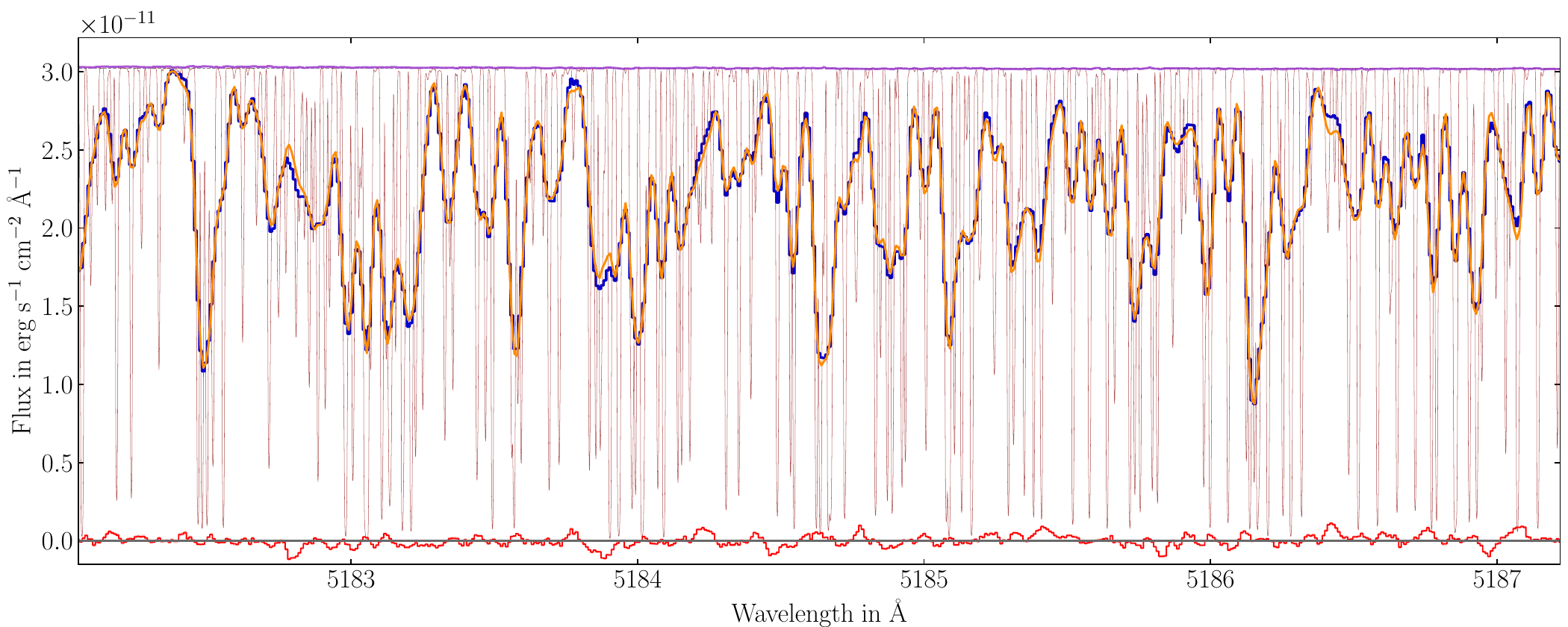}%
 \caption{One chunk of the spectrum visualizing the fitting procedure described in Section~\ref{Sec:FitProcedure}. Data shown as dark blue histogram corresponds to the observed spectrum of HD\,69081 taken through the iodine cell. The pure stellar continuum obtained through an empty cell is shown at the top in purple.
 The shifted and scaled theoretical iodine spectrum is shown as thin line in the background and the convolved forward model fitted to the data in orange. Differences between model and data are indicated in red at the bottom.
 Uncertainties of observed quantities are in principle shown as shaded areas around the histograms but in practice barely exceed the adopted line thickness and are therefore mostly invisible.}
 \label{Fig:FitI2}
\end{figure*}

The fit of model to the observed data by optimization of the model parameters $\mathBF{f_0}$, $\Boldsymbol{\tau}$, and $\mathBF{\Delta{}v}$ is performed using a standard $\chi^2$-minimizer.
The adopted variance vector is based on the propagated uncertainties of the observed iodine absorption spectrum, provided by the data reduction procedure, and the (fully negligible) uncertainties in the composite of the stellar spectrum obtained through the empty cell.
The fitting process is visualized in Figure~\ref{Fig:FitI2}. The shown spectral range corresponds to one full block over which the fit is performed.
The dark blue histogram represents the observed $\II$ absorption spectrum. The purple, featureless continuum indicates the stellar spectrum observed through the clear cell.
In thin lines, the plot shows the intrinsic iodine absorption spectrum as provided by the model (see Section~\ref{Sec:I2Model}), which is thermally broadened but apart from this kept at infinite spectral resolution.
Therefore, it exhibits a plethora of narrow absorption lines and showcases that every iodine absorption feature in the ESPRESSO spectrum is a blend of numerous individual  $\II$ transitions.
This also highlights the great importance of the instrumental LSF, which strongly influences the amount of blending and, given the high dynamic range in the intrinsic $\II$ spectrum ranging from weak to almost saturated lines, leads to a non-trivial change of the observed $\II$ feature shape when re-scaling the optical depth.
Thus, an accurate model of the LSF is indispensable for the forward modeling of the observed absorption spectrum.
Based on the visual impression, this forward model, shown with orange color in Figure~\ref{Fig:FitI2}, describes the observed absorption spectrum in general quite well. Residuals are indicated by the red curve at the bottom of the plot.
Considering the residuals and the formal uncertainties, one can conclude that the shown example is formally not a good fit. The residuals are obviously not consistent with the statistical uncertainties but dominated by systematics. This is not particularly surprising, given the extremely high S/N ($\lesssim \VAL{400}\:\textup{per}\:\sqrt{\kmps}$) of the data and the few parameters optimized during the fit.
Most discrepancies do not arise from an imperfect scaling of the model, neither in intensity, optical depth, or shift in wavelength direction, but instead are related to small, localized regions in which the $\II$ model simply contains too much or too little absorption.
One example for this is the region around $\VAL{5186.3}\,\mathrm{\AAA}$, clearly exhibiting less absorption than predicted by the model.
These residuals might just be related to imperfections in the $\II$ model.
The description of the iodine potential levels presented by \citet{Knöckel2004} is fairly complex, yet it still contains only about \VAL{250} model parameters to predict millions of $\II$ absorption lines over a wide spectral range. Given this, it provides a fairly accurate description of the observations.
Also, we do not fit for the actual temperature of the absorption cell during operations. Instead, we use a model that was calculated for a fixed temperature of $\VAL{60}^\circ\textup{C}$, which is probably close to the real value but certainly not entirely accurate. Since the relative intensity of different branches of transitions is temperature dependent, a certain variation of line strengths is expected. A possible future improvement of the analysis would therefore be to consider models with a range of cell temperatures and infer this property during the fit instead of assuming it.
However, considering that the goal of our analysis is to extract wavelength information and not a perfect modeling of the $\II$ spectrum, we consider the quality of the fit shown in Figure~\ref{Fig:FitI2} good enough for now to proceed.

\section{Validation of accuracy on sky}
\label{Sec:AccuracySky}

To validate the accuracy of the ESPRESSO wavelength calibration, we observed the bright, featureless, and fast rotating star HD\,69081 (spectral type B2, $V \approx 5\,\mathrm{mag}$) through the iodine cell. The data was analyzed as described in Section~\ref{Sec:Analysis} by fitting a forward model of the iodine absorption to the observed spectrum.
Here, the $\II$ model by \citet{Knöckel2004} provides a reliable energy scale only for wavelengths $\gtrsim\VAL{5150}\,\AAA$.  At the blue end, this somewhat limits the spectral range over which the analysis can be performed. The iodine spectrum itself contains usable lines down to nearly $\VAL{5000}\,\AAA$. Towards longer wavelengths, the iodine absorption becomes progressively weaker and the obtained results noisier. We thus only include data up to $\lambda < \VAL{6350}\,\AAA$.
Nevertheless, this covers \VAL{22} of the 78 echelle orders of ESPRESSO, each fitted with up to \VAL{16} blocks and in two slices, resulting in nearly \VAL{700} inferred sets of parameters $\left(\, \mathBF{f_0}, \; \mathBF{\color{black}\Delta{}v}, \; \Boldsymbol{\tau} \,\right)$ per spectrum.

\begin{figure*}
 \includegraphics[width=\linewidth]{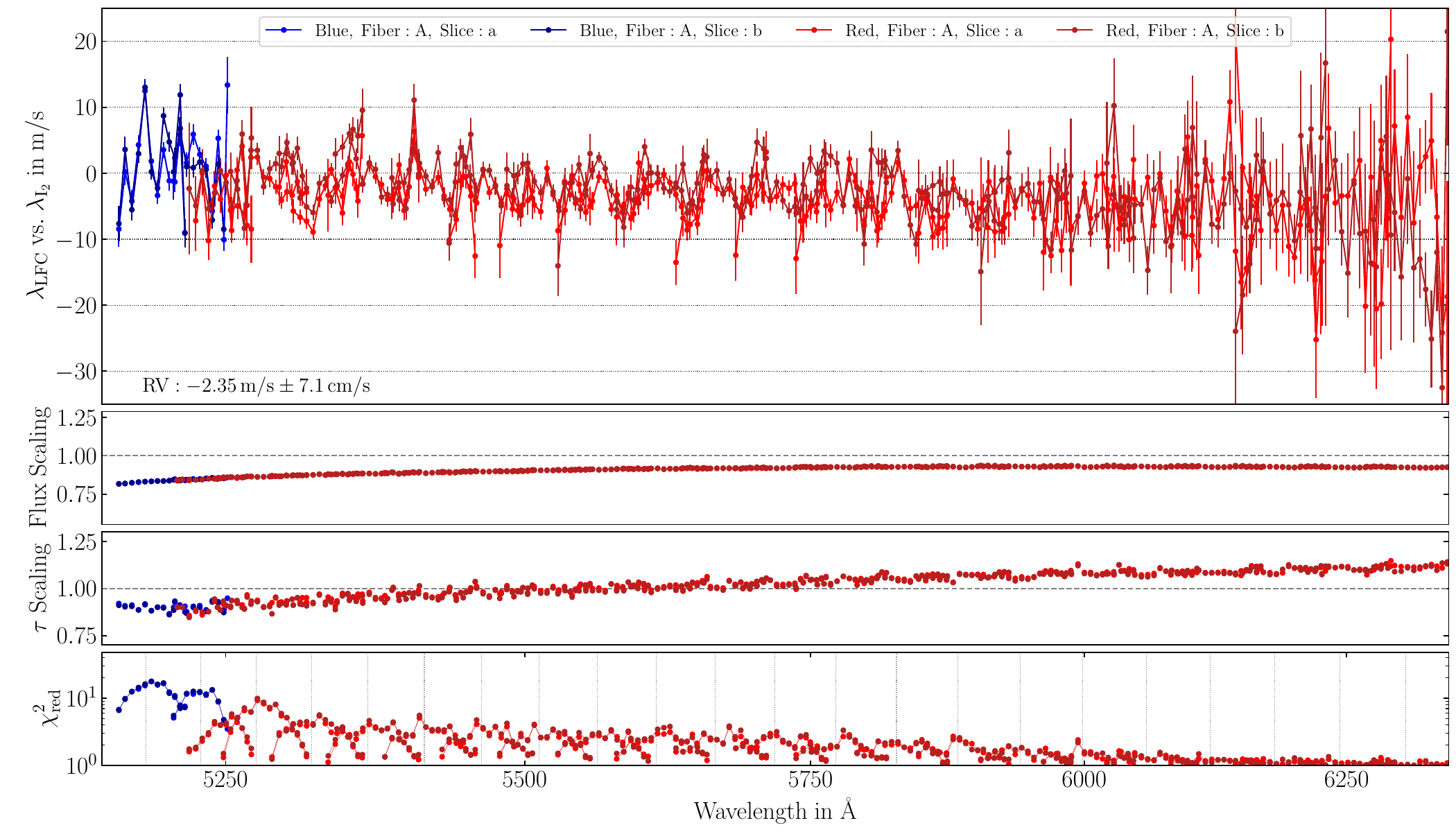}
 \caption{
 On-sky test of the wavelength calibration accuracy.
 The top panel shows the velocity difference between the LFC-based wavelength solution of ESPRESSO and the FTS-calibrated wavelength scale of the $\II$ model, inferred following the fit procedure described in Section~\ref{Sec:FitProcedure}. The other panels indicate the applied scaling in flux and optical depth of the iodine absorption, as well as the reduced $\chi^2$ value. Each datapoint in the plot corresponds to one spectral chunk, which were all fitted independently. Colors indicate data obtained from the different spectrograph arms and slices. Due to the two spectral slices and overlap between orders, up to four measurements might be present at a given wavelength. The central wavelength of each order is indicated in the bottom panel by a vertical dotted line. The global RV offset between the $\II$ spectrum and the LFC wavelength solution is shown in the bottom-left corner of the top panel.}
 \label{Fig:I2_Accuracy}
\end{figure*}

The results for an observation taken in twilight on May 23, 2023, are shown in Figure~\ref{Fig:I2_Accuracy}.
The exposure was integrated for $\VAL{105}\,\mathrm{s}$ and peak photon counts reached about \VAL{25\,k}~ADU per pixel, i.e. slightly less than half the saturation threshold of the detector in $\texttt{1HR2x1}$ binning mode.
In consequence, the data has a high signal-to-noise, reaching 
up to $S/N \approx \VAL{400}\:\textup{per}\:\sqrt{\kmps}$ in each of the two spectral slices. This leads to typical uncertainties in $\Delta{}v$ between $\VAL{1}$ and $\VAL{2}\,\mps$ for each individual block and a total RV content of about $\VAL{7}\,\cmps$.

The lower panels of Figure~\ref{Fig:I2_Accuracy} show the resulting $\chi^2_\mathrm{red}$ and the re-scaling factors for the $\II$ optical depth and overall flux.
The distribution of $\chi^2_\mathrm{red}$ follows the blaze function and is therefore strongly correlated with the order structure of the spectrograph. This demonstrates again, as already shown in Figure~\ref{Fig:FitI2}, that the residuals between model and data are dominated by systematic effects and higher S/N in the center of the orders simply leads to worse $\chi^2_\mathrm{red}$. One can also notice that neighboring spectral orders overlap substantially in wavelength.
The overall trend of $\chi^2_\mathrm{red}$ with wavelength, i.e. worsening towards shorter wavelengths is most-likely simply correlated with the strength of the iodine absorption, which is strongest at shorter wavelengths and gradually fades-out towards the red.
The absolute value of the $\Boldsymbol{\tau}$~rescaling has no particular meaning, since the $\II$ model was calculated for an arbitrary column density.
One notices however, that there is a trend with wavelength. The optical depth has to be scaled-down to about $\VAL{90}\,\%$ towards the blue end of the wavelength range but scaled up to $\VAL{\approx 115}\,\%$ at the long-wavelength end.
The flux calibration of the observations taken through the iodine absorption cell and the ones with a clear cell, taken on subsequent evenings, are relatively consistent and require only a mild, but still chromatic re-scaling of the intensities.

The actual quantity of interest, however, is displayed in the top panel of Figure~\ref{Fig:I2_Accuracy}. It shows the determined velocity offset, $\mathBF{\color{black}\Delta{}v}$, between the wavelength scale of the iodine model and the wavelength calibration of the spectrograph and therefore provides an assessment of the wavelength calibration \textit{accuracy}.
For this comparison, the LFC wavelength solution is preferred over the ThAr/FP one, since it is supposed to provide a better \textit{accuracy}.
As displayed in Figure~\ref{Fig:I2_Accuracy}, excellent agreement between the two wavelength scales at the level of a few $\mps$ is achieved.
Also, the two spectrograph slices provide fully consistent results, emphasizing the high quality of the data reduction, LSF modeling, and wavelength calibration procedure. The assumption of a Gaussian LSF, for comparison, would lead to discrepancies at the level of tens of $\mps$.
Nevertheless, one can spot some regions and spectral orders where the difference between the two  slices significantly exceeds the formal uncertainty.
This is not entirely surprising, given that systematic offsets between the slices of up to $\pm1\,\mps$ have already been reported by \citet{Schmidt2024}, for instance in the comparison of the ThAr lines to the LFC wavelength solution.
In addition, the bluemost measurement in nearly each echelle order has the tendency to exhibit a negative deviation. Similarly, but less pronounced, several redmost blocks exhibit a positive deviation. This behavior is most-likely related to an imperfect model of the instrumental LSF. As lined out in \citet{Schmidt2024}, the ESPRESSO LSF shape tends to change quite rapidly towards the edges of each echelle order, in particular the blue ends. The LFC, despite containing several hundred thousand lines, does not provide entirely sufficient information to model this change of the LSF with fully satisfactory precision and accuracy. To reach a higher fidelity on the LSF model, one would require a wavelength-tunable calibration source. By taking multiple exposures with different offsets of the line positions, one would obtain a substantially denser sampling of the instrumental LSF and could therefore model it in smaller spectral chunks. Tunability of the ESPRESSO LFC is clearly intended and desired, but unfortunately still not available in practice.

Apart from these systematics, which are most-likely related to the spectrograph and the data processing, one also notices in the top panel of Figure~\ref{Fig:I2_Accuracy} an additional and relatively fast modulation of the residuals that does not correlate with the blaze function of the spectrograph and is mostly consistent among the two slices. The origin of this is therefore probably related to the modeling of the iodine spectrum. Some rapidly oscillating modulation that correlates with the $\II$ absorption band structure has been seen by \citet{Reiners2024} and should in principle be corrected by the supercalibration process. There might be the possibility that this correction does not entirely remove the effect, however, the amplitude reported by \citet{Reiners2024}, about $\pm1\,\mps$, seems actually too small to explain the modulation found in the ESPRESSO observations. Further investigations are therefore necessary to pinpoint and correct the origin of these systematics.

On large spectral scales, $\II$ and LFC wavelength scale are fairly consistent. Averaging all measurements, the formal offset is just $\VAL{-2}\,\mps$, However, the scatter shown in the top panel of Figure~\ref{Fig:I2_Accuracy} is clearly dominated by systematics, so the actual result of the averaging will depend on the exact selection of data points.
For instance, there is some indication for a mild wavelength dependence in the $\mathBF{\color{black}\Delta{}v}$ measurements, exhibiting a slope of maybe $\VAL{-5}\,\mps$ over the $\approx\VAL{1000}\,\AAA$ analyzed here, although such a global trend can easily be confused with the other, more local, systematics present at a similar level.

We stress that our data reduction of the ESPRESSO spectra does not introduce any long-range correlation. For the LFC wavelength solution, the longest coherence scale is substantially shorter than an echelle order. All spectral elements further apart than this are treated fully independently \citep{Schmidt2021, Schmidt2024}.
In the super-calibration of the $\II$ model, \citet{Reiners2024} report some slope in their wavelength calibration of $\approx\VAL{50}\,\cmps$ per $\VAL{1000}\,\AAA$, but judge this not significant due to systematics. Still, some of the slope noticeable in Figure~\ref{Fig:I2_Accuracy} might be related to the calibration of the $\II$ wavelength scale.
Also, the absolute calibration of the $\II$ model was performed around $\VAL{8250}\,\AAA$, exploiting the linearity of the FTS wavelength scale. Therefore, a slope\,--\,if present\,--\,could also contribute to the total offset of $\VAL{2}\,\mps$.

Despite these numerous small systematics, the consistency between $\II$ and LFC at the level of a few $\mps$ is indeed remarkable, in particular given the vastly different nature of the two type of spectra. This confirms and validates the accuracy of the LFC wavelength solution as well as the iodine wavelength scale at the stated level.
Given this, one can be highly confident that the same wavelength calibration accuracy can also be reached on actual science observations.
Noteworthy is as well that the global offset between $\II$ and LFC is found to be substantially smaller than for LFC and ThAr/FP solution. For these, one finds a mean offset of $\VAL{-8}\,\mps$ and a spread from about $\VAL{-12}\,\mps$ to $\VAL{+2}\,\mps$, as shown in \citet{Schmidt2024}.
The good agreement between $\II$ and LFC provides further evidence that the discrepancy between LFC and ThAr/FP wavelength solution is indeed  related to the thorium laboratory wavelengths \citep{Redman2014}.
Beyond this, the obtained $\II$ observations are of excellent quality and provide the required reference data to identify and pinpoint systematics at the $1\,\mps$ accuracy level, which is a crucial first step for further improvements of the ESPRESSO wavelength calibration procedures.

\section{Accuracy in daytime observations}
\label{Sec:AccuracyFlat}

To obtain more extended time series observations than possible in twilight alone, additional absorption spectra were taken during daytime. For this, a halogen lamp and a diffuser were placed slightly in front of the absorption cell, providing illumination similar to classical flatfield observations.
In this setup, the light obviously did not pass through the telescope and it entered the $\II$~cell and spectrograph front-end with an unusual beam geometry, but this allowed to take data independently of other operations at the observatory and thus facilitated continuous $\II$ time-series observations over several hours.

\begin{figure*}
 \includegraphics[width=\linewidth]{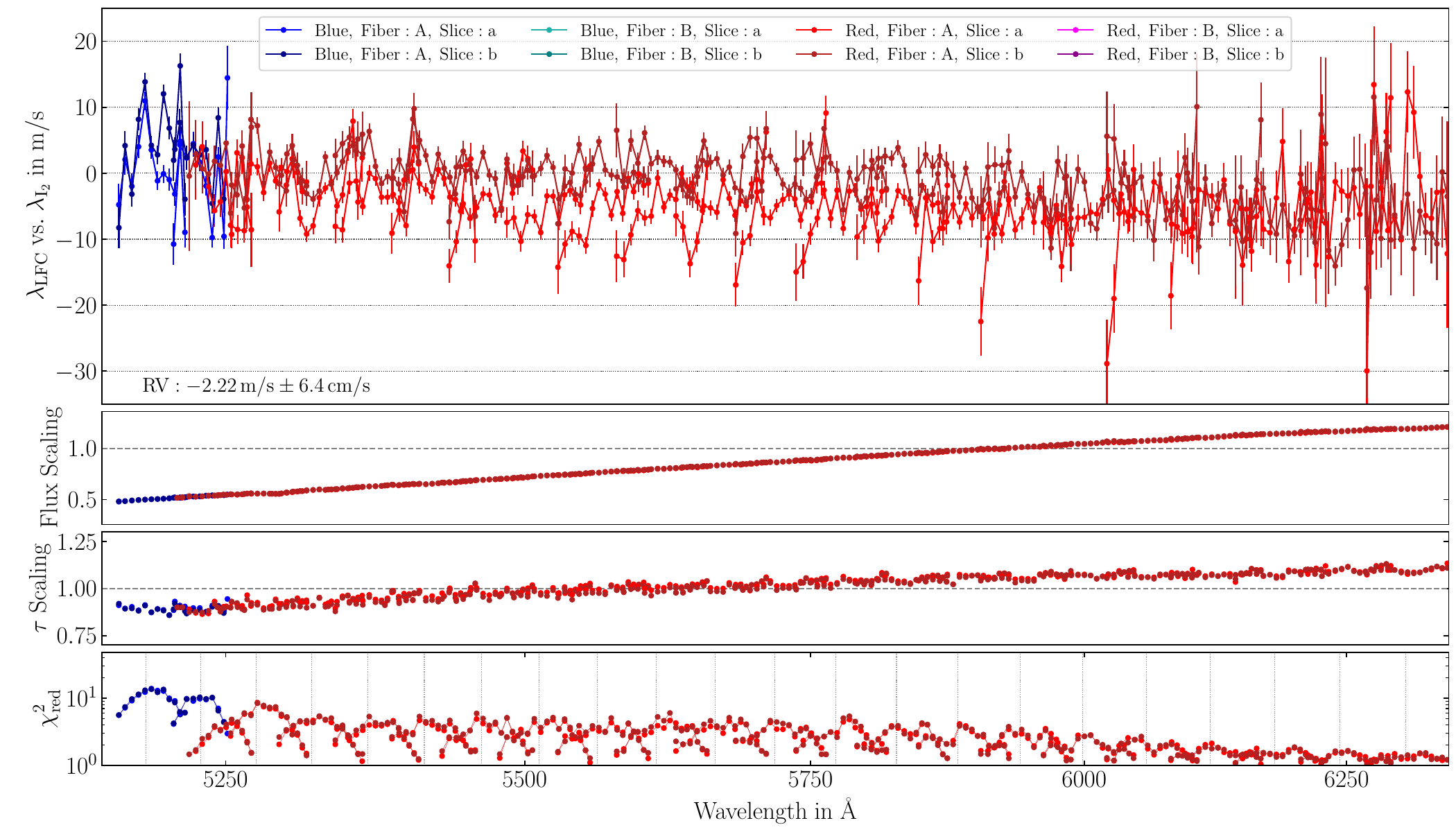}%
 \caption{Assessment of the wavelength calibration accuracy from daytime flatfield exposures. The analysis is identical to the one shown in Figure~\ref{Fig:I2_Accuracy}, but based on spectra taken during the day by illuminating the absorption cell with a lamp instead of using stellar light fed from the telescope.}
 \label{Fig:I2_Accuracy_Flats}
\end{figure*}

The data were reduced and processed identically to the one described in the previous section.
The only exception is that no spectra through an empty cell were taken. Instead, a first-order polynomial (instead of just a scaling factor) was fitted for each block to directly describe the continuum. This is fully adequate since the spectrum of the lamp is completely featureless and the spectral size of the blocks small.
The result is shown in Figure~\ref{Fig:I2_Accuracy_Flats}.
Obviously, the $\chi^2_\mathrm{red}$ as well as the  $\Boldsymbol{\tau}$-scaling show very similar distributions as in Figure~\ref{Fig:I2_Accuracy}
and the flux scaling naturally differs due to the different approach for the continuum treatment. The velocity offset between $\II$ and LFC wavelength scale, however, exhibits some striking peculiarities.
From $\VAL{5300}$ to $\VAL{5900}\,\mathrm{\AAA}$, the velocity offsets measured in the two slices are not consistent anymore as it was for the stellar observations. Instead, $\Delta{}v$ in Slice~a is on average about $5\,\mps$ below the one observed in Slice~b.
Also, the outliers at the blue ends of many echelle orders are substantially more pronounced. All in all, the fits to the daytime observations are not as clean and tidy as the stellar observations and exhibit much more scatter and systematics.

The data in Figure~\ref{Fig:I2_Accuracy} and \ref{Fig:I2_Accuracy_Flats} were taken on different days (May 23 and May 25), but both were reduced and calibrated, including the LSF, using the standard calibration exposures of the corresponding morning. Also, at no point during the analysis of the full dataset, encompassing nearly two weeks, were significant changes of the instrument noticed. Taking data on a different day should therefore have no impact.
The only relevant difference between the two datasets lies in the illumination of the iodine cell and therefore different beam geometry.
The telescope delivers for the twilight observations a slowly converging beam ($\approx\,f/22$) to the ESPRESSO front-end. In the ad-hoc daytime setup, the beam geometry formed by the diffuse and unfocused  lamp is poorly characterized but certainly diverging and should completely overfill the spectrograph fiber. The image of a star, however, follows a Moffat profile \citep{Moffat1969, Schmidt2024b} and\,--\,at least under reasonable seeing conditions\,--\,is significantly concentrated towards the center of the 1" fiber.
This leads to a different population of modes in the spectrograph fiber. Of course, care has been taken to minimize the susceptibility of ESPRESSO to the details of the fiber-injection geometry, e.g. by the introduction of a \textit{double-scrambler} \citep{Hunter1992} in the fiber train and the use of octagonal fibers.
Still, the obvious presence of additional systematics in Figure~\ref{Fig:I2_Accuracy_Flats} and in particular their correlation with the echelle structure, strongly suggest that mode scrambling is not perfect and the different illumination geometries of the spectrograph fiber indeed lead to significantly different instrumental LSFs. Under such circumstances, the LSF determined from the LFC calibration frames is not representative anymore for the LSF encountered in the daytime $\II$ frames 
and thus leads to inaccurate results.

The beam geometry produced by the lamp is probably far off from the one delivered by the telescope, but the observed systematics at a level of $\VAL{5\,\mps}$ also fairly significant, in particular considering the ambitious wavelength calibration goals of ANDES.
While the effect reported here results from the unusual beam geometry produced by the lamp, a similar difference exists in principle also between the light coming from the calibration unit and the telescope (see Figure~\ref{Fig:Sketch1}).
Their beam geometries might be similar but certainly not identical and given that fiber-injection geometry obviously matters, one has to ask how representative the LSF derived from calibration sources is for observations taken through the telescope.
ESPRESSO follows the approach to simply overfill the spectrograph fiber with calibration light. There is no attempt to emulate the telescope pupil, e.g. by using a mask that casts a central obstruction representing the secondary mirror of the telescope.
The critical aspect is that a possible misalignment of the light injection from the calibration unit, or its drift with time, can hardly be monitored or measured. The expected cosmological signal for the redshift drift experiment carried out with ANDES amounts to only a few $\cmps$ per decade and there might remain a concern that a spurious contribution to that measurement might be caused by a changing fiber-injection geometry for light coming from the calibration unit.
The validation with an absorption cell, however, always probes exactly the same light path as the science exposures.
It therefore provides additional information that cannot be acquired from the calibration unit and is therefore ideal to validate the wavelength calibration established from spectra of the calibration sources. Regular monitoring of the wavelength calibration by means of absorption cell observations might therefore be crucial to build confidence in the validity of the scientific measurements, in particular concerning science cases as ambitious as the redshift drift experiment.

\section{Short-term Stability}
\label{Sec:StabilityShort}

\begin{figure}
 \includegraphics[width=\linewidth]{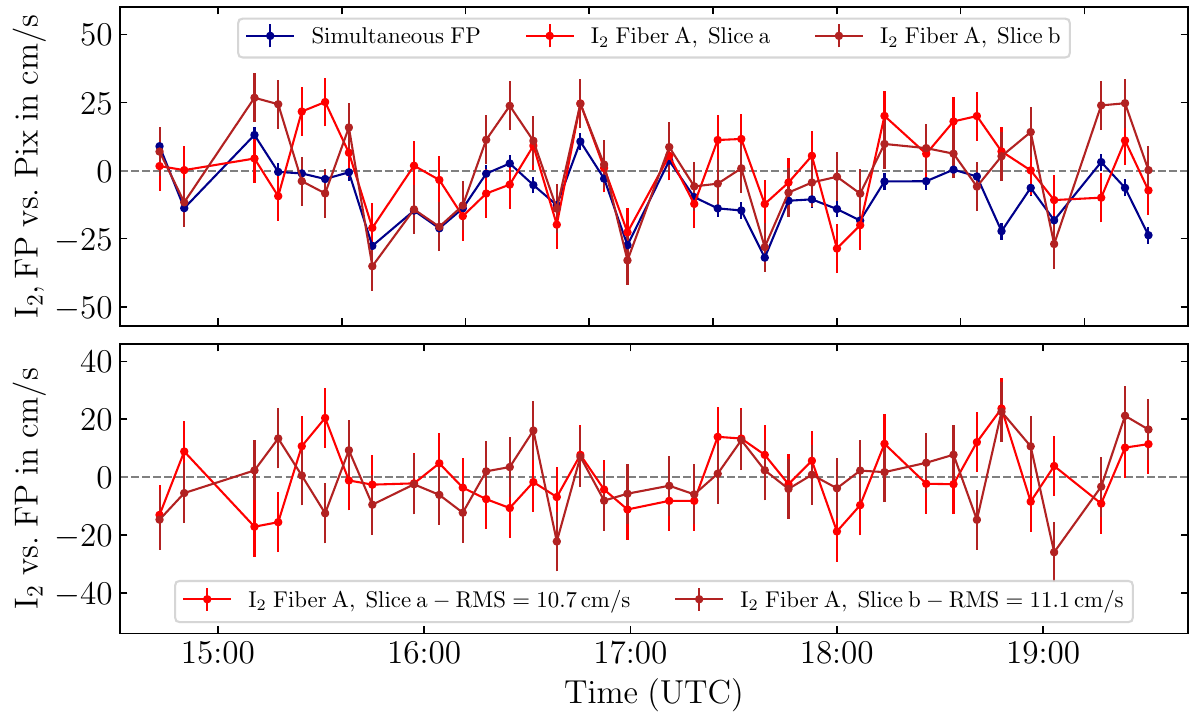}%
 \caption{Short-time stability of the ESPRESSO wavelength calibration inferred from daytime exposures. The upper panel shows the measured raw drift of the iodine spectrum in both spectrograph slices and of the simultaneous FP w.r.t. the detector pixel coordinates over a course of about \VAL{five} hours. Since most of this drift is actually related to the spectrograph itself, the simultaneous FP was used to correct the instrumental drift, as it is done for science observations. The bottom panel therefore shows the $\II$ measurements against the drift-corrected wavelength scale. Errorbars indicate the uncertainty based on propagated photon noise and after averaging measurements over the full spectral range. The determined RMS scatter of the measurements is stated in the legend.}
 \label{Fig:I2_Stability_Short}
\end{figure}

Although the wavelength information derived from the daytime observations might be \textit{inaccurate}, the observed systematics should nevertheless remain virtually unchanged over limited time periods and thus still be useful to assess the \textit{stability} of the setup. To do so, a sequence of \VAL{37} daytime $\II$ observation taken over a period of \VAL{five hours} was analyzed.
The result is presented in Figure~\ref{Fig:I2_Stability_Short}.
The top panel contains for both slices the change of the globally averaged velocity offsets, $\mathBF{\color{black}\Delta{}v}$, between $\II$ and ThAr/FP wavelength solution. The latter one was derived from the morning calibrations and kept static throughout the sequence, i.e. no drift corrections were applied. The plot therefore shows raw drifts with respect to the detector coordinates. The instrumental drift derived from the simultaneous FP in Fiber~B is instead shown explicitly. Obviously, $\II$ and FP drifts track each other quite well, indicating a common origin, most-likely caused by a movement of the ESPRESSO detectors corresponding to approximately $\pm20\,\cmps$.

In the bottom panel of Figure~\ref{Fig:I2_Stability_Short}, the simultaneous FP was used to correct for the instrumental drift, leading to calibrated $\II$ drift measurements, fully equivalent to the procedure used for science observations.
One can notice that the data points here are not simply the difference between raw $\II$ and FP drifts displayed in the top panel. The reason is that globally-averaged quantities are shown in Figure~\ref{Fig:I2_Stability_Short}. The drift corrections derived from the simultaneous FP spectra are, however, applied on a order-by-order basis.
The calibrated $\II$ drifts obtained in this way show a remarkable stability. They are in general very close to zero, mostly consistent for both slices, and exhibit a RMS scatter (over time) of about \VAL{11\,\cmps} for each of the two slices.
The statistical uncertainty for individual measurements, estimated from propagated photon noise, is approximately $\VAL{10\,\cmps}$, indicating that the inferred calibrated drift measurements are very close to the photon-noise limit with a fairly low contribution by additional systematics or jitter.
It demonstrates the outstanding stability of ESPRESSO and the quality of the verification that can be reached using $\II$ absorption cells on astronomical spectrographs, perfectly in line with the laboratory tests presented by \citet{Debus2023} and \citet{Reiners2024}.

\section{Long-term stability}
\label{Sec:StabilityLong}

Although impressive, the short-term test presented above is not fully representative for the RV stability one could achieve in actual science observations.
Instead of just a few hours, these extend over weeks or even years and require a regular, e.g. daily, re-calibration of the instrument. The test presented in Figure~\ref{Fig:I2_Stability_Short}, however, exploits the short-term stability of ESPRESSO, e.g. in terms of the LSF, the overall shape of the echellogram, and the general raw RV stability. This means that a static ThAr/FP wavelength solution could be used and the small residual intra-day drifts tracked and efficiently corrected using the simultaneous FP, which itself is also extremely stable to better than $2\,\cmps$ over one day.
For longer time scales, however, one cannot rely in this way on the stability of the setup. Instead, it is necessary to fully re-calibrate the instrument, typically on a daily basis. This requires numerous extra steps (order localization, trace-profile determination, flatfielding, LSF characterization, FP $D_\mathrm{eff}(\lambda)$ determination, establishment of wavelength solutions, ...) which all introduce a certain amount of noise and thus lead to reduced \textit{precision}, but are of course necessary to capture occurring changes in the instrument and guarantee \textit{accuracy}.
The achievable long-term stability will therefore always be worse than the much simpler short-term stability.
Its quality will depend on the additional statistical noise introduced by the numerous calibration steps and how well systematic changes in the instrument are captured by the instrument model.

To perform a comprehensive test of the long-term stability, we use iodine absorption cell observations of HD\,69081 taken during six consecutive evening twilights. The extent of this sequence was limited by the overall duration of the $\II$ experiment and barely qualifies as \textit{long-term}, however, data for every day are reduced and calibrated in a fully independent way. From every set of morning calibrations, a completely new instrument model is constructed, including a full characterization of the LSF, and then used to calibrate the observations taken during twilight. There is no memory in these calibrations from one day to the next and no mechanism that exploits the inherent stability of the instrument.
Therefore, measurements from the six days are fully independent and it should actually not matter when these observations were taken, i.e. on consecutive evenings or months apart.
The presented test can thus be expected to be well representative for the long-term RV stability achievable with ESPRESSO.

\begin{figure}
 \includegraphics[width=\linewidth]{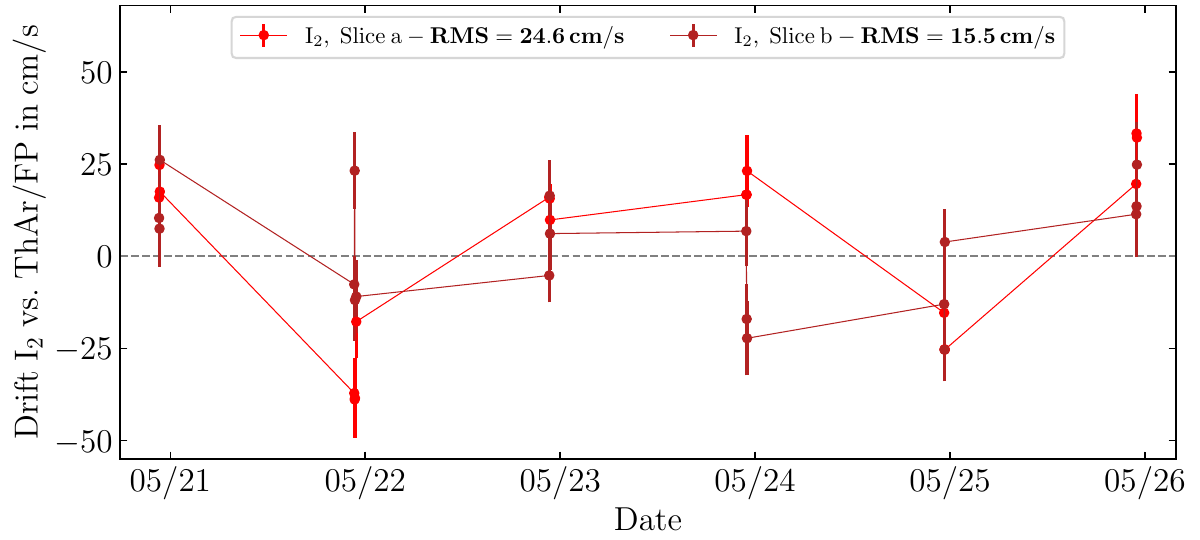}%
\caption{On-sky test of the ESPRESSO wavelength calibration stability derived from iodine absorption cell observations of HD\,69081 over six days. The drift is given against the ThAr/FP wavelength solution which was corrected for intra-day drifts using the simultaneous FP calibration. Observations from each evening twilight were reduced and calibrated using the standard calibrations from that corresponding morning. This includes a full re-determination of the instrumental LSF. Measurements from consecutive days are therefore fully independent and make no assumption about stability of the instrument. Although covering only six days, this test is therefore representative for the long-term wavelength stability of ESPRESSO. Errorbars indicate the uncertainty based on propagated photon noise and after averaging measurements over the full spectral range. The determined RMS scatter of the measurements is stated in the legend.}
 \label{Fig:I2_Stability_Long}
\end{figure}

Figure~\ref{Fig:I2_Stability_Long} shows the $\II$ drift measurements, typically three per evening, obtained during the six days.
The presented analysis embodies a full end-to-end test, from sky to final RV measurement, completely analogous to actual science observations, with the only exception that the analyzed absorption features are not of astrophysical origin but imprinted by the iodine cell.
Since no suitable astrophysical source exists on sky, this must be considered the closest and most representative end-to-end validation of the instrumental RV stability.
We stress here that the reported drifts are given w.r.t. the ThAr/FP wavelength calibration. It turns out that it provides more \textit{stable} measurements than the independent LFC calibration. The reason behind this are rapid fluctuations in the line-to-background flux ratio in the LFC spectra. Even for consecutive exposures, this changes significantly in some spectral regions and\,--\,due to unavoidable imperfections in the data reduction\,--\,leads to substantial variations in the derived wavelength solution by more than $\VAL{1\,\mps}$ in some orders, which is in excess of the stability reachable by the $\II$ observations. Given the currently available data processing tools, these flux instabilities in the LFC spectra therefore limit the achievable RV stability.
The FP, on the other hand, exhibits very little variations in flux and the ThAr/FP scheme therefore provides a more \textit{stable} (but not as \textit{accurate}) wavelength solution compared to the LFC.

\begin{figure}
 \includegraphics[width=\linewidth]{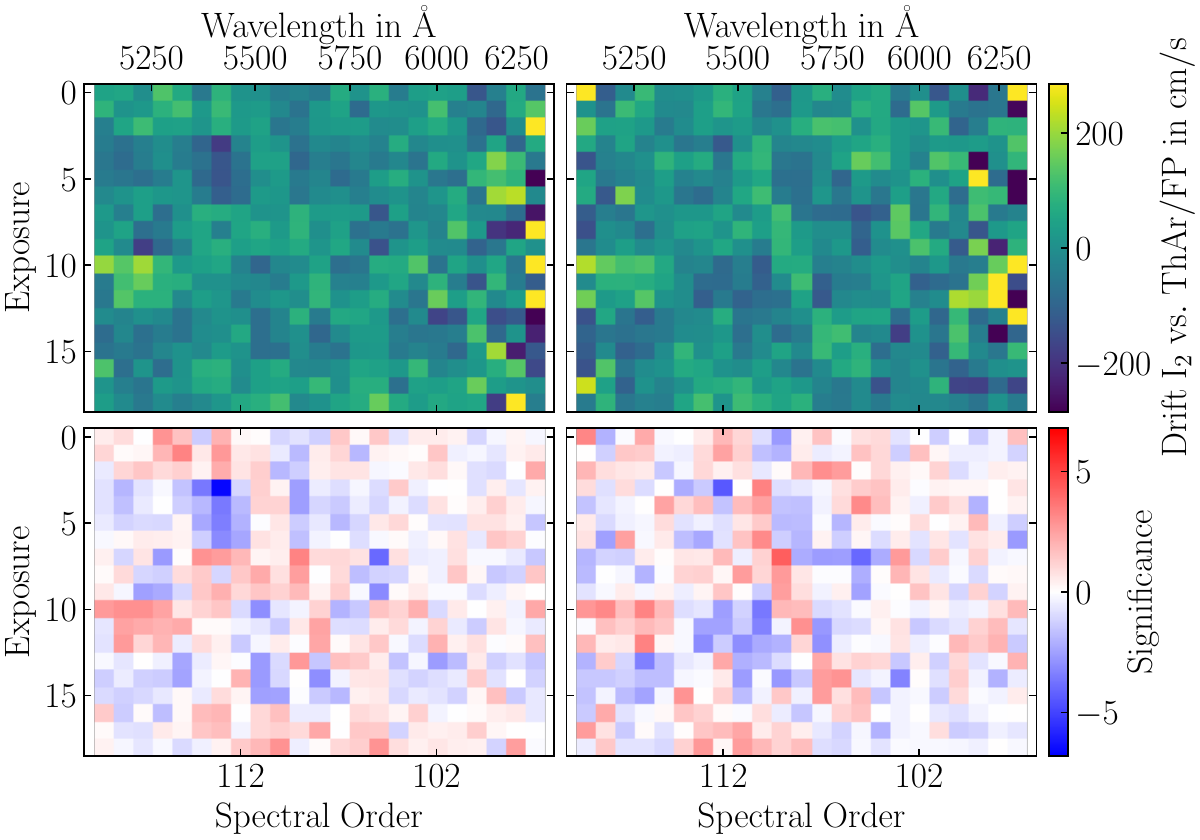}%
\caption{Same data as in Figure~\ref{Fig:I2_Stability_Long} but displayed for individual spectral orders. Rows in the panels correspond to different exposures taken over the six days, while columns correspond to the different orders of the spectrograph and therefore wavelengths. The left panels visualize the measurements derived from Slice~a, while the right ones for Slice~b. The top panels show the inferred drift of the $\II$ spectrum w.r.t. \VAL{the ThAr/FP wavelength solution}. In the bottom panels, this is normalized by the statistical uncertainty and therefore shows the formal significance of the drift.}
 \label{Fig:I2_Stability_Long_Orders}
\end{figure}

As indicated in Figure~\ref{Fig:I2_Stability_Long}, we find for Slice~b a RMS scatter over all 19 exposures of just $\approx\VAL{15.5\,\cmps}$. This is\,--\,as expected\,--\,slightly larger than for the short-term stability presented in Figure~\ref{Fig:I2_Stability_Short} and also in excess of the purely statistical photon noise of about $\VAL{10\,\cmps}$, but nevertheless an excellent result, given that it involves a daily re-calibration of the instrument model including a re-determination of the LSF. Therefore, a similar level of stability should also be achievable over much longer time scales.
For Slice~a, however, we find a substantially larger scatter of almost $\approx\VAL{25\,\cmps}$. The outliers here are the measurements from the evening of the 21st. These deviate coherently by $\approx\VAL{30\,\cmps}$ compared to Slice~b.
The high information content of the $\II$ spectra allows to conduct the stability analysis on an order-by-order basis and separately for both slices, which is shown in Figure~\ref{Fig:I2_Stability_Long_Orders}.
One notices for Slice~a of the observations taken on May~21 (i.e. exposures 3 to 6) a cluster of drift measurements in spectral Orders \VAL{114 to 112} which deviate.
In particular for spectral Order~\VAL{113} one finds a drift of $\VAL{-2.9\,\mps}$, corresponding to a significance of $\VAL{6.6}\,\sigma$. No other outlier with a significance $\ge5\,\sigma$ is found anywhere else in the dataset and in particular not in Slice~b. Excluding the observations from May 21 or the spectral orders \VAL{114 to 112} would lead to a RMS scatter comparable to Slice~b.

We investigated further and could determine that the excess scatter in fact roots in the calibrations taken on the morning of the 21st and not in the actual $\II$ observations. Reducing the absorption cell exposures from the 21st with the calibrations either from the 20th or 22nd substantially reduces the deviation of the Slice~a measurements and leads to a RMS scatter over the full six days of $\approx\VAL{18\,\cmps}$, only marginally larger than the excellent stability we find for Slice~b.

Despite careful investigation, we were not able to identify the ultimate cause for this issue. Numerous aspects can be excluded though.
Following our calibration scheme, there are only few effects that can affect the two slices differentially. For instance, the calibration of the FP effective gap size, $D_\mathrm{eff}(\lambda)$, based on the ThAr exposures, is applied globally for all fibers and slices, because it is an inherent property of the FP device.
Also all other aspects related to the calibration sources should affect both spectrograph slices identically.
The instrumental LSF, of course, is unique for every slice and fiber. We closely investigated this aspect and found apart from the unavoidable noise in the process no evidence that the LSF models derived from the calibrations of the 21st would in any way be outliers or significantly deviant from the ones found on the other days.

The simultaneous drift correction was initially applied on an order-by-order basis and individually for each slice, which should be the most accurate but also more noisy approach. Since there is no strong evidence and also no plausible reason why drifts should be different for individual orders and slices, we re-reduced the data applying global drift corrections for each detector, which also simplifies the whole process. This, however, did not lead to a substantial change regarding the outliers in Slice~a on May 21st.
We also switched for the simultaneous FP spectra from the \textit{optimal extraction} algorithm \citep{Horne1986, Zechmeister2014} to direct summation of the counts.
Given that \textit{optimal extraction} is not optimal \citep[see e.g.][]{Schmidt2021}, this indeed lead to intra-day drifts which are different from the ones determined from the optimally extracted spectra and clearly indicates that there are systematics present at the level of tens of $\cmps$. However applying any of the methods consistently to the full dataset always lead to more or less the same RMS scatter and in particular always showed the excursion of the measurements from the 21st.

We also directly compared line positions in the FP wavelength calibration frames (labeled \texttt{WAVE,FP,FP} in the ESPRESSO data flow) from the six consecutive days and confirmed that there were no large drifts of the spectrograph over the course of the experiment. The issue therefore needs to come from the calculation of the wavelength solution.
Given the presence of the \textit{beat pattern noise} \citep{Schmidt2021}, which leads to an additional scatter of individual FP line positions beyond photon noise by up to $8\,\mps$, and the rather flexible Gaussian-process approach adopted \citep{Schmidt2024}, the intra-order wavelength solutions are for a given pixel position inherently noisy by up to $\pm1\,\mps$. This flexible approach was chosen to guarantee best possible \textit{accuracy} of the wavelength solution but is not necessarily optimal for stability. To achieve this, one could decide to fit a substantially stiffer model for the intra-order wavelength solution. Ensuring improved \textit{accuracy} and \textit{stability}, however, would require to overcome the issue causing the \textit{beat pattern noise} and by this drastically reduce the effective uncertainty associated with each individual line position measurement.

Although we were in the end not able to correct this issue, it clearly showcases the potential of the iodine absorption cell technique and its ability to identify and characterize such systematics with very high precision, i.e. at the level of tens of $\cmps$ for the whole spectrum or $\mps$ on scales of individual orders.
In addition, the stability achieved on Slice~b, or when excluding the data from the 21st, i.e. at the level of $\VAL{15}$ to $\VAL{18\,\mps}$, represents the most-stringent end-to-end demonstration of RV stability conducted so far with an astronomical spectrograph.
For comparison, the exoplanet with the smallest RV amplitude found so far is Proxima~Cen~d \citep{Faria2022}. It causes a reflex motion of its host star of only $39\,\cmps$ and required a highly sophisticated Gaussian-process modeling of the stellar activity to become detectable.
\citet{Faria2022} also explicitly report in their analysis a residual {instrumental jitter} term of $27\,\mathrm{cm/s}$, already indicating the remarkable performance of ESPRESSO, but still larger than the stability demonstrated here.
Nevertheless, the found systematics also showcases that further improvement to the wavelength calibration process are necessary, to break the $10\,\cmps$ RV barrier and reach the goal of discovering a twin Earth, but even more in the context of ANDES and the redshift drift experiment.

We have so far emphasized that the $\II$ absorption spectra are highly representative for actual science observations.
This is entirely true with respect to the optical light path and the spectral shape.
However, science observations would in addition be affected by the variable barycentric velocity correction. This effect can of course not be emulated by absorption cells. In the experiment, the $\II$ lines did in all exposures always fall on the same pixels of the detector. Our analysis, however, did not explicitly exploit this fact and we therefore do not consider it a limitation.
The adopted spectral chunks are quite broad and contain many $\II$ lines. Also do the features of the iodine spectrum have no special relation to the calibration lines (ThAr, FP, or LFC) used to establish the wavelength solutions. A shift of the $\II$ spectrum should therefore not cause a strong systematic effect.
We stress, however, that\,--\,due to the variable barycentric correction\,--\,wavelength calibration \textit{accuracy} is still highly important even when only caring for \textit{stable} RV measurements. Any gradient (with respect to wavelength or velocity) in the wavelength calibration accuracy would directly translate to a systematic, periodic modulation (in time) of the inferred radial velocities.
Fortunately, we were able to demonstrate that our wavelength calibration is rather accurate (see Figure~\ref{Fig:I2_Accuracy}) and without any strong global gradient.
The numerous small-scale modulations noticeable in Figure~\ref{Fig:I2_Accuracy} might contribute, but they should at least in parts be related to an imperfect match of the $\II$ model and not actually a distortion of the wavelength solution. In addition, such small-scale distortions would for sources with a large number of absorption features distributed over a considerable range in wavelengths, like the $\II$ cells or the Ly$\alpha$ forest, average down substantially.
Nevertheless, we stress that any spectrograph aiming for precise RV measurements has to be equipped with a capable exposure meter, required to accurately determine the flux-weighted mean barycentric correction \citep{Tronsgaard2019}.
Due to the chromaticity of numerous effects, e.g. atmospheric extinction and fiber-injection losses, this has to be done independently in numerous spectral channels across the wavelength range and therefore requires a chromatic exposure meter \citep[see e.g.][]{Landoni2014,Blackman2017,Blackman2019}.

\section{Complementary Analysis}
\label{Sec:FlucGradientAnalysis}

The forward-modeling approach used in the previous sections requires knowledge of the intrinsic iodine cell spectrum and the instrumental line profile. It has the advantage of yielding the absolute wavelength difference between the iodine cell and the LFC or ThAr/FP wavelength solutions -- i.e.\ the results shown in Figure~\ref{Fig:I2_Accuracy} -- and is capable of capturing changes of the LSF which might occur on longer time scales. However, a measurement of stability alone is also possible with a model-free approach: the spectra of the fast-rotating star taken through the iodine cell can be directly compared to each other to derive the wavelength shift between them. This will complement, and offer a cross-check, on the results in Figure~\ref{Fig:I2_Stability_Long} from the forward-modeling approach. In this section we briefly describe our implementation of this complementary analysis.

\subsection{Data reduction and analysis}\label{ss:8_analysis}

Here we use the same exposures of HD\:69081 taken through the $\II$ cell over 6 days used in Section~\ref{Sec:StabilityLong}. To ensure this alternative analysis is independent of the forward-modeling results, we reduced these exposures with different software -- the standard ESO data reduction pipeline for ESPRESSO \citep[][version 3.1.1]{Pepe2021}. For each exposure, two extracted spectra were produced of each slice, calibrated with the two different standard day-time exposures, i.e.\ the ThAr/FP  and LFC exposures. These were typically taken \VAL{1--2} and \VAL{9--10}\,h before the star exposures, respectively.

Direct comparison of the observed iodine spectra is possible because the velocity shifts between them, on the detector itself, are much smaller than a spectral pixel: the $\II$ absorption occurs in the observatory rest-frame, so there is no barycentric velocity variation between exposures over different days. Also, ESPRESSO is highly stabilized and instrumental drifts are very small ($\VAL{\lesssim1\,\mps}$ per day) and the LSF remains virtually unchanged over the duration of this experiment. This allows us to apply the `flux gradient' method of \citet{Bouchy2001} to obtain the velocity shift between two spectra. We treat \VAL{6} exposures taken on the last day of the campaign (UT date \VAL{2023-05-25}) as a `reference set', combining their extracted spectra to form a higher S/N `reference' spectrum. Before combination, these spectra were normalized by the combination of \VAL{3} exposures of HD\,69081, taken on the same night, without the iodine cell. The flux array in each slice, in each order, was then scaled so that the median values of all \VAL{6} $\II$ exposures matched, and the combined `reference' was formed from the weighted mean (scaled) flux.

Each $\II$ exposure from the \VAL{6} days was then compared to this higher S/N reference spectrum using the flux gradient method generalized to the case where the reference has non-negligible (albeit significantly lower) statistical noise \citep{Liske2008}. Each slice, of each order, of a `target' spectrum was first normalized by a combined HD\,69081 spectrum without $\II$ absorption and then broken into chunks of \VAL{200} spectral pixels ($\approx$\VAL{100}\,\kmps, or \VAL{1.8}\,\AA\ at 5500\,\AA). The reference spectrum was re-dispersed (i.e.\ its flux was re-sampled) onto the wavelength grid of the target chunk \citep[using {\sc spectres},][]{Carnall2017}, and its flux array was scaled so that its median value matched that of the target chunk. The velocity shift between the target and reference chunk, and its 1$\sigma$ uncertainty, was then calculated using equations~10--13 of \citet{Liske2008}, where the flux gradient at pixel $i$ was determined via a parabolic fit to its flux and that of its neighboring two pixels.

\subsection{Results}\label{ss:8_results}

\begin{figure}
\begin{center}
\includegraphics[width=0.95\columnwidth]{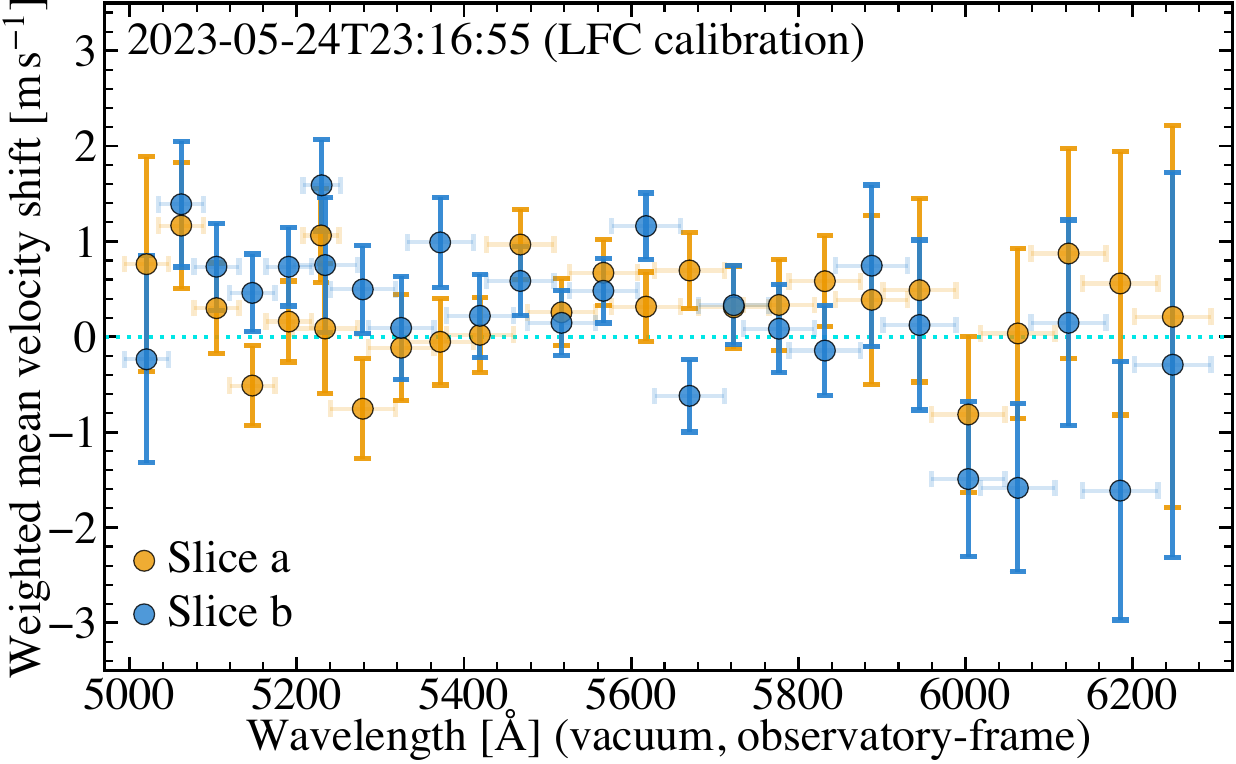}\vspace{-0.5em}
\caption{Velocity shift between echelle orders of a single target $\II$ exposure (date and time as labeled) and the reference set taken $\approx$1 day later. The value for each order is the weighted mean from non-overlapping 200-pixel chunks ($\approx$100\,\kmps, or 1.8\,\AA\ at 5500\,\AA) across the order, with 1$\sigma$ uncertainties plotted.}
\label{Fig:dv_allords_MTM}
\end{center}
\end{figure}

Figure \ref{Fig:dv_allords_MTM} shows the weighted mean velocity shift, for each echelle order with $\II$ absorption, between a single target exposure (on \VAL{2023-05-24}) and the reference set (the following day). We did not observe significant variation in the velocity shift across echelle orders (not plotted), so taking the weighted mean across each order provides meaningful data points here. In this example, the target exposure's wavelength scale was calibrated using the LFC and the velocity shift is less than \VAL{2\,\mps} across the $\II$ wavelength range. In this case, there is little-to-no evidence of systematic shifts between the two slices. However, this is not true of all exposures when using LFC calibrations, but it is true when using ThAr/FP calibrations (see below). Finally, in Figure~\ref{Fig:dv_allords_MTM} there is no systematic shift between the blue and red arms (there are two data points for each slice where the two arms overlap at $\sim$5200\,\AA), and this is true generally for the exposures from all \VAL{6}~days as well.

\begin{figure*}
\begin{center}
\centerline{\hbox{
    \includegraphics[height=0.27\textheight]{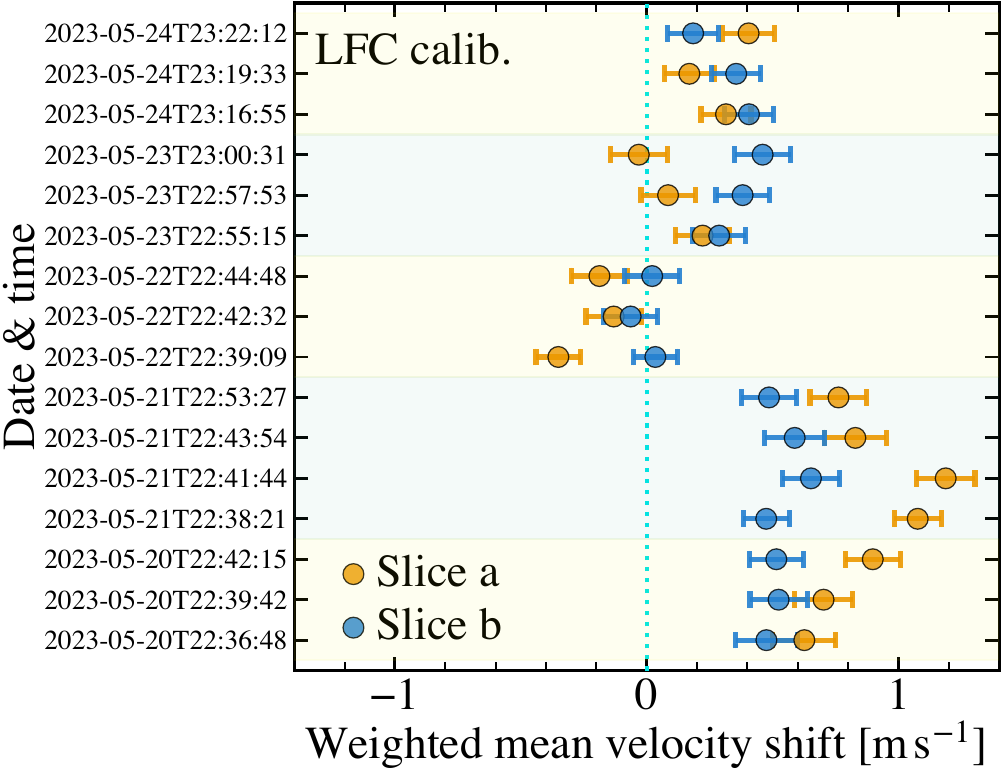}
    \hspace{0.01\columnwidth}
    \includegraphics[height=0.27\textheight]{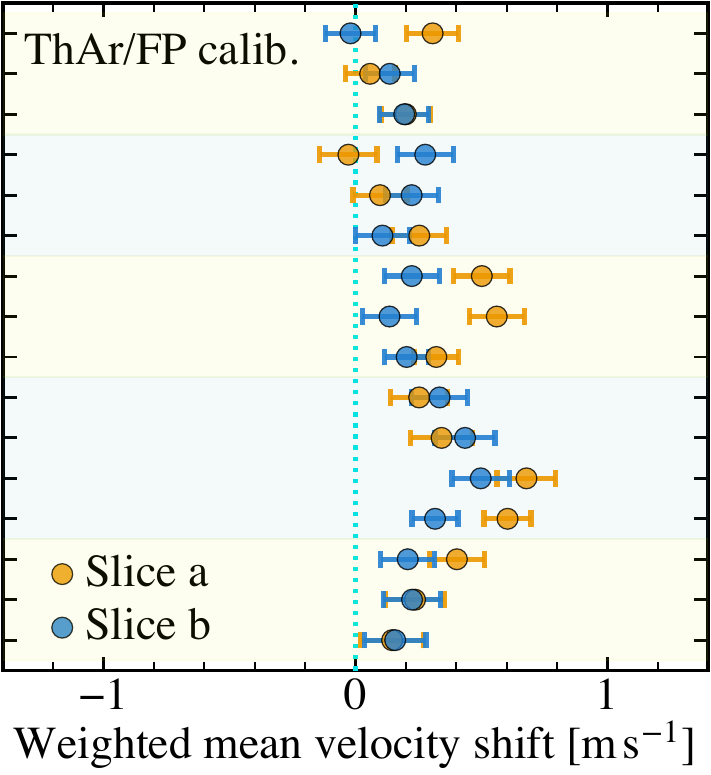}
}}
\caption{Weighted mean velocity shift and 1$\sigma$ uncertainty for each on-sky I$_2$ exposure, relative to the reference set (2023-05-25), using both LFC calibration (left panel) and ThAr/FP calibration (right). The vertical axis shows the date and time of the exposure, with the light yellow and green shading in the panels delineating different days. Zero velocity shift from the reference is indicated by the vertical dotted line in each panel.}
\label{Fig:dv_vs_date_MTM}
\end{center}
\end{figure*}

Figure \ref{Fig:dv_vs_date_MTM} shows the velocity shift (averaged over all chunks in all orders) between the reference set and each target exposure in the \VAL{5}~days beforehand. It also compares the results obtained when calibrating the target and reference exposures with the LFC and ThAr/FP exposures. Focusing first on the ThAr/FP results (right-hand panel), we note the general consistency of the results over the \VAL{5}~days, especially for Slice~b which exhibits an offset of $\VAL{22\pm3\,\cmps}$ (weighted mean) from the reference set. Indeed, its RMS variation of $\VAL{12\,\cmps}$\ is only marginally larger than the mean uncertainty, $\VAL{11\,\cmps}$. The results for Slice~a scatter slightly more than expected from purely random, with an RMS of $\VAL{20\,\cmps}$ and the same mean uncertainty ($\chi^2$ per degree of freedom, $\nu$, is $\chi^2=3.5$). Nevertheless, they generally agree very well with Slice~b's shift values. Overall, these results indicate that the on-sky iodine cell observations can reliably track drifts in the spectrograph and/or the ThAr/FP calibration down to the $\VAL{\sim20}\,\cmps$ level, and that the spectrograph and calibration are stable at that level over multi-day timescales.

In the left panel of Figure~\ref{Fig:dv_vs_date_MTM} we see a contrasting situation for the LFC-calibrated versions of the same spectra. The velocity shifts for the \VAL{2}~days preceding the reference set are reasonably consistent with the reference, and are similar to the results from the corresponding ThAr/FP calibrated spectra. However, the results from the \VAL{3}~days before those show marked differences to the ThAr/FP ones, and are not consistent with each other. Overall, the LFC results for Slice~a and b plotted in Figure~\ref{Fig:dv_vs_date_MTM} have RMS variations of $\VAL{20}$ and $\VAL{46\,\cmps}$, with the same mean uncertainties as before (i.e.\ $\VAL{11\,\cmps}$). The results for Slice~a scatter substantially more than Slice~b, just like for the ThAr/FP results. The remarkable differences between the LFC and ThAr/FP results shows, as stated before, that the LFC wavelength solution is less stable than the ThAr/FP calibration.

This complementary analysis fully confirms the results found in the previous sections.
Both methods demonstrate the excellent stability of ESPRESSO and the high precision to which this can be tested and verified using iodine absorption cells, either by the full forward-modeling approach or the direct comparison of the spectra using the flux gradient method.

\section{Summary and Conclusions}

Key science drivers of the future ANDES spectrograph, in particular the search for a possible variation of fundamental physical constants and the redshift drift experiment, impose requirements on the quality of the wavelength calibration far beyond the current state of the art. The goal here is to achieve an \textit{accuracy} of $1\,\mps$ and \textit{stability} of $1\,\cmps$ over the lifetime of the instrument.
Demonstrating that these demands are fulfilled and the inferred scientific results are indeed genuine and not affected by instrument systematics requires an independent validation of the wavelength calibration.
This has to probe the full optical path, from sky to detector, and include the data processing, identical to actual science observations, because differences in the light injection geometry between calibration unit and telescope can cause minute but significant differences in the inferred wavelengths.
Unfortunately, no astrophysical source exists on sky that would provide reference spectra with even remotely the required stability or accuracy.

We therefore propose to use absorption cells for validation of the wavelength calibration.
The strategy is to observe a bright and featureless star through an absorption cell that imprints the highly stable and well known spectral features of iodine \citep{Knöckel2004, Reiners2024}. This setup therefore emulates the spectrum of an ideal on-sky reference, which can be analyzed like any other science observation and thus reveals possible systematics in the wavelength calibration of the spectrograph or the data processing.

To demonstrate this approach in practice, we temporarily installed an iodine cell at ESPRESSO in May 2023 and conducted numerous tests with it.
Observing the bright B-type star HD\,69081 during twilight through the absorption cell, we were able to demonstrate wavelength calibration \textit{accuracy} at the level of a \VAL{few $\mps$} by comparing the calibrated wavelength scale of the iodine spectral model to the ESPRESSO LFC wavelength solution (see Section~\ref{Sec:AccuracySky}).

Illuminating the cell with a lamp revealed a much poorer agreement and significant systematics.
Obviously, the instrumental LSF produced by the unusual illumination of the spectrograph fiber differed significantly from the one in the LFC spectra, which are used to establish the LSF model for the analysis. This mismatch led to deviations in the wavelength scale up to $\VAL{\sim5\,\mps}$ (Section~\ref{Sec:AccuracyFlat}). It demonstrates that the measured wavelengths of spectral features do indeed depend on the light injection geometry, despite the use of octagonal fibers and double-scramblers.
A similar mismatch of the instrumental LSF can in principle also occur between the calibration spectra provided by the calibration unit and the science spectra observed through the telescope. This difference can with current methods not be calibrated and a slow change of the fiber injection geometry over decades might induce a spurious signal in the redshift drift experiment. This emphasizes the need for a validation procedure that probes the full optical path, identical to actual science observations.

Furthermore, a test of short-term \textit{stability} revealed an RMS scatter between the $\II$ wavelength scale and the drift-corrected ThAr/FP wavelength solution of just $\VAL{12\,\cmps}$ over the course of \VAL{five hours} (Section~\ref{Sec:StabilityShort}).
More importantly, stability of $\VAL{16\,\cmps}$ was demonstrated, at least in Slice~b, over the full duration of the measurement campaign, i.e. \VAL{six days} (Section~\ref{Sec:StabilityLong}). Here, observations from each evening twilight were reduced and calibrated fully independently, based on the default instrument calibrations of the corresponding morning.
Therefore, no assumptions about intrinsic instrument stability were made and the analysis, even though lasting only for six days, should thus be fully representative for the long-term stability of ESPRESSO measurements.
The found scatter of the measurements in Slice~a is substantially larger, reaching nearly $\VAL{25\,\cmps}$.
However, the $\II$ spectra allowed to determine that the issue must have its origin in the calibration frames taken on May 21st and is mostly limited to a few spectral orders, highlighting the great diagnostic capabilities of the iodine spectra.
In addition, we were able to confirm these excellent levels of stability with a fully complementary analysis, using a different software package for data reduction and the flux gradient method instead of a forward-modeling of the spectrum (Section~\ref{Sec:FlucGradientAnalysis}).
This $\II$ absorption cell experiment therefore represent the most stringent end-to-end test of wavelength calibration \textit{accuracy} and \textit{stability} conducted with any extreme-precision RV spectrograph so far.
Given the nature of the produced spectra, i.e. utilizing an on-sky source and imprinting a dense forest of absorption features, the test is highly representative for actual science observations and demonstrates the current capabilities of ESPRESSO.
Here, the presented experiment shall be understood as a proof-of-concept study. The absorption cell technique bears extensive potential for future monitoring and improvement of ESPRESSO, but might also become an integral part of the ANDES wavelength calibration strategy.

There are numerous aspects in which this method can be further developed and improved. For instance, we have presented good reasons why the conducted six day test should indeed be representative for the long-term instrument stability.
However, it is of course highly desirable to obtain a much more extensive dataset and actually demonstrate this in practice.
Ideal would be a continuous monitoring campaign that regularly, e.g. once per week, acquires spectra taken through the absorption cell. Such observation could be conducted during twilight and as part of the general instrument calibration plan, very similar to the observation of spectrophotometric standard stars.
However, making this feasible requires full automatization of the process.  Manually placing the cell into the beam was fine given the temporary nature of the experiment. For regular monitoring, however, it would be necessary to place the cells on a motorized stage and fully include their operation in the instrument control software, allowing to command absorption cell observations through the standard ESO observing block templates. Provided this, a regular $\II$ monitoring campaign would pose a limited and well justifiable extension of the existing instrument calibration plan and not cause excessive additional operational efforts or require any valuable night time. The main challenge here lies in the adaptation of the ESPRESSO instrument control system.

On the data analysis side, a clear next step would be to infer the gas temperature of the iodine cell, instead of a~priori assuming one.
The underlying $\II$ model can in principle be computed for various temperatures. It is therefore possible to keep the gas temperature a free parameter in the inference process. This might very well lead to a better fit of the model to the data, e.g. in terms of $\chi^2$ metric, as well as a reduction of the residual systematics noticeable for instance in Figure~\ref{Fig:I2_Accuracy}.

Some of these discrepancies in the the comparison between $\II$ wavelength scale and the spectrpgraph wavelength solution, however, are unrelated to the $\II$ spectra and appear already during the wavelength calibration process, namely the slight offset between slices by about $\VAL{\pm1\,\mps}$ \citep{Schmidt2024}. The most likely cause for this are inaccuracies in the determination of the instrumental LSF.
As outlined already in \citet{Schmidt2024}, the LFC provides a vast number of lines, but these are nevertheless just barely sufficient to derive a truly accurate model of the instrumental LSF.
Also the increase in the scatter between the short-term and long-term stability test from $\VAL{12\,\cmps}$ to $\VAL{16\,\cmps}$ (and even more for Slice~a, see Figure~\ref{Fig:I2_Stability_Short} and \ref{Fig:I2_Stability_Long}) indicates that the full instrument calibration procedure adds additional noise. This is neither surprising nor avoidable, but it would of course be desirable to keep this noise as small as possible. The dominant contributor is most likely also the determination of the LSF.
The exact amount could be studies and quantified in detail, but the available data already suggests that better and more informative calibration spectra are necessary to significantly improve this aspect of the instrument calibration.
Getting a larger number of independent lines and a drastic reduction of sampling noise, however, can only be achieved by taking multiple calibration exposures with different line positions.
It would therefore be highly desirable to finally establish the long-foreseen tuning capability of the ESPRESSO LFC.
This would facilitate a more accurate determination of the LSF  and most-likely reduce the instrument-related systematics visible in Figure~\ref{Fig:I2_Accuracy}, e.g. the excursions at the blue and red ends of nearly each spectral order and the discrepancy between slices.
However, it is understood that frequency tuning of a LFC is a rather complex procedure and might be in conflict with other design goals. For future spectrographs like ANDES, one might therefore consider to
develop a dedicated LSF calibration source, e.g. based on a relatively high-finesse FP etalon, that does not need to provide the extreme accuracy or stability of existing LFC or FP calibrators, but provides many narrow lines with a~priori known shape and is easily tunable.

Another issue of the ESPRESSO LFC is the high and unstable diffuse background \citep{Schmidt2024}.
The  new type of photonic-crystal fiber installed in May 2022 has substantially extended the available spectral range, now covering wavelengths down to $\VAL{4300\,\AAA}$. However, this comes at the expense of drastically reduced flux, strong temporal variations in the intensity of the lines and especially fluctuations of the line-to-background ratio from one exposure to the next.
Accurately subtracting the diffuse LFC background is challenging and rarely perfect, in particular when the wings of the lines start to overlap and when no a~priori characterization of the laser itself is available. This currently limits the fidelity of the derived LSF model. The changing line-to-background flux ratio, on the other hand, has a negative impact on the stability of the wavelength calibration. Over extended sequences taken as part of this measurement campaign, variations from exposure to exposure by more than $\VAL{1\,\mps}$ in certain spectral orders were noticed. Here, the data was already analyzed using the accurate non-parametric LSF model. Assuming a Gaussian LSF increases the instability to over $\VAL{4\,\mps}$.
For this reason, the test for instrument stability presented in Section~\ref{Sec:StabilityLong} was performed against the ThAr/FP solution, which is in practice substantially more stable than the LFC solution. The same was found in our complementary analysis, revealing a substantially higher scatter when utilizing the LFC wavelength solution (Figure~\ref{Fig:dv_vs_date_MTM}).
These aspects currently hamper progress on the ESPRESSO wavelength calibration and highlight that further technical development is necessary,
in particular regarding the still immature LFC technology, but probably also exploring alternatives like dedicated FP-based LSF calibrators.

A quite different aspect concerns the simultaneous drift correction.
For this study, we used the simple but highly efficient gradient method presented by \citet{Bouchy2001} to measure the shift of the FP spectrum relative to the pixel coordinates of the detectors. However, the assumptions this algorithm makes about the spectrum, i.e. that it does not change except by a small displacement in spectral direction, might not be entirely satisfied. For instance, there is some crosstalk between spectrograph fibers, reaching levels up to $0.5\%$, and it has been noticed that the simultaneous drift measurement from one fiber exhibits some spurious drift depending on which type of source is injected into the other fiber. Also, apparent differential shifts between the two spectrograph slices have been observed, which as well appear not to be real but rather an artifact of contamination and imperfect data processing. These effects are small, e.g. a few tens of $\cmps$, but become relevant at the extreme level of RV precision we aim for.
Improvements to the quality of the simultaneous drift correction might be possible by using a more sophisticated, but also computationally more expensive algorithm, for instance by explicitly fitting each individual FP line.
The important point here is that the iodine spectra provide a sufficiently clean and precise ($\VAL{\approx10\,\cmps}$ per exposure) reference dataset, free from other sources of noise like stellar jitter, and therefore allow to identify, characterize, and ultimately fix such issues.

One general shortcoming of iodine absorption cells, however, is the limited spectral range they cover, providing absorption lines only from about $\VAL{5000}$ to $\VAL{6300\,\mathrm{\AAA}}$. End-to-end verification of the wavelength solution in this region is certainly very informative and provides unique insights not obtainable from the sources in the calibration unit alone.
However, it would of course be highly desirable to have verification over the full wavelength range used for scientific measurements.
Here, tellurium absorption cells might provide a viable path forward. As demonstrated recently by \citet{Ross2022a, Ross2022b}, molecular tellurium provides a very dense forest of lines, quite similar to the $\II$ absorption spectrum, from at least $\VAL{4100}$ to $\VAL{5300\,\mathrm{\AAA}}$.
One complication and probably the reason why tellurium has so far not been used in astronomy is that the cells have to be operated at a temperature of about $\VAL{620^\circ\mathrm{C}}$.
This seems problematic at first, however, commercially available bench-top furnaces allow at least for laboratory setups easy and uncomplicated operation of tellurium absorption cells at these temperatures. It would certainly require further studies of the $\Te$ spectrum and some engineering to adapt such cell furnaces for use at astronomical observatories, but this does not appear entirely unfeasible.  Thus, molecular tellurium might indeed be a viable option to extend the absorption cell method to substantially larger wavelength ranges. The combination of $\Te$ and $\II$ absorption cells would provide continuous coverage from $\VAL{4100}$ to $\VAL{6300\,\AAA}$, which represents the dominant part of the the wavelength range relevant for the redshift drift experiment.
These could in the IR complemented by cells filled with organic compounds \citep[e.g.][]{Mahadevan2009, Anglada-Escude2012}.

As demonstrated in this study, iodine absorption cells provide a simple, easy, and practicable solution for end-to-end validation of the wavelength calibration of high-resolution spectrographs. By observing bright and featureless stars through the absorption cell, one emulates a science target with precisely known spectral features and therefore probes the full optical path from sky to the detector.
Using this, we were able to demonstrate wavelength calibration accuracy and stability of ESPRESSO at an unprecedented level. We strongly suggest to permanently install an absorption cell at ESPRESSO and use it for a continuous monitoring of the instrument.
We also advocate to include absorption cells in the ANDES design to provide validation of the wavelength calibration alongside observations taken for the redshift drift experiment or the search for a possible variation of fundamental physical constants.

\section*{Acknowledgements}

The described iodine absorption cell experiment was financially supported by the Big Questions Institute (BQI), Sydney.
TMS would like to thank Fran\c{c}ois Bouchy for extensive support under the SNF synergia grant CRSII5-193689 (BLUVES).
MTM acknowledges the support of the Australian Research Council through \textsl{Future Fellowship} grant FT180100194 and the Australian Research Council Centre of Excellence in Optical Microcombs for Breakthrough Science (project number CE230100006) which is funded by the Australian Government.
The work of CJM was financed by Portuguese funds through FCT (Funda\c c\~ao para a Ci\^encia e a Tecnologia) in the framework of the project 2022.04048.PTDC (Phi in the Sky, DOI 10.54499/2022.04048.PTDC). CJM also acknowledges FCT and POCH/FSE (EC) support through Investigador FCT Contract 2021.01214.CEECIND/CP1658/CT0001 (DOI 10.54499/2021.01214.CEECIND/CP1658/CT0001).
PH would like to thank the Federal Ministry of Education and Research for the funding of the project: Verbundprojekt 05A2023  ELT -ANDES: Design und Konstruktion des ANDES Spektrographen für das Extremely Large Telescope (ELT).
This work has been carried out within the framework of the National Centre of Competence in Research PlanetS supported by the Swiss National Science Foundation.
This research has made use of Astropy, a community-developed core Python package for Astronomy \citep{Astropy2013,Astropy2018,Astropy2022}, and Matplotlib \citep{Hunter2007}.

\section*{Data Availability}

All data used for this study are publicly available from the ESO archive facility under program id 60.A-9680(A):\\\url{http://archive.eso.org/cms/data-portal.html}.


\bibliographystyle{mnras}
\bibliography{../Literature}

\newcommand{\noop}[1]{}
\begin{thebibliography}{}
\makeatletter
\relax
\def\mn@urlcharsother{\let\do\@makeother \do\$\do\&\do\#\do\^\do\_\do\%\do\~}
\def\mn@doi{\begingroup\mn@urlcharsother \@ifnextchar [ {\mn@doi@}
  {\mn@doi@[]}}
\def\mn@doi@[#1]#2{\def\@tempa{#1}\ifx\@tempa\@empty \href
  {http://dx.doi.org/#2} {doi:#2}\else \href {http://dx.doi.org/#2} {#1}\fi
  \endgroup}
\def\mn@eprint#1#2{\mn@eprint@#1:#2::\@nil}
\def\mn@eprint@arXiv#1{\href {http://arxiv.org/abs/#1} {{\tt arXiv:#1}}}
\def\mn@eprint@dblp#1{\href {http://dblp.uni-trier.de/rec/bibtex/#1.xml}
  {dblp:#1}}
\def\mn@eprint@#1:#2:#3:#4\@nil{\def\@tempa {#1}\def\@tempb {#2}\def\@tempc
  {#3}\ifx \@tempc \@empty \let \@tempc \@tempb \let \@tempb \@tempa \fi \ifx
  \@tempb \@empty \def\@tempb {arXiv}\fi \@ifundefined
  {mn@eprint@\@tempb}{\@tempb:\@tempc}{\expandafter \expandafter \csname
  mn@eprint@\@tempb\endcsname \expandafter{\@tempc}}}

\bibitem[\protect\citeauthoryear{{Anglada-Escud{\'e}}
  et~al.,}{{Anglada-Escud{\'e}} et~al.}{2012}]{Anglada-Escude2012}
{Anglada-Escud{\'e}} G.,  et~al., 2012, \mn@doi [\pasp] {10.1086/666489}, \href
  {https://ui.adsabs.harvard.edu/abs/2012PASP..124..586A} {124, 586}

\bibitem[\protect\citeauthoryear{{Astropy Collaboration} et~al.,}{{Astropy
  Collaboration} et~al.}{2013}]{Astropy2013}
{Astropy Collaboration} et~al., 2013, \mn@doi [\aap]
  {10.1051/0004-6361/201322068}, \href
  {https://ui.adsabs.harvard.edu/abs/2013A&A...558A..33A} {558, A33}

\bibitem[\protect\citeauthoryear{{Astropy Collaboration} et~al.,}{{Astropy
  Collaboration} et~al.}{2018}]{Astropy2018}
{Astropy Collaboration} et~al., 2018, \mn@doi [\aj] {10.3847/1538-3881/aabc4f},
  \href {https://ui.adsabs.harvard.edu/abs/2018AJ....156..123A} {156, 123}

\bibitem[\protect\citeauthoryear{{Astropy Collaboration} et~al.,}{{Astropy
  Collaboration} et~al.}{2022}]{Astropy2022}
{Astropy Collaboration} et~al., 2022, \mn@doi [\apj]
  {10.3847/1538-4357/ac7c74}, \href
  {https://ui.adsabs.harvard.edu/abs/2022ApJ...935..167A} {935, 167}

\bibitem[\protect\citeauthoryear{{Baranne}, {Mayor}  \& {Poncet}}{{Baranne}
  et~al.}{1979}]{Baranne1979}
{Baranne} A.,  {Mayor} M.,   {Poncet} J.~L.,  1979, \mn@doi [Vistas in
  Astronomy] {10.1016/0083-6656(79)90016-3}, \href
  {https://ui.adsabs.harvard.edu/abs/1979VA.....23..279B} {23, 279}

\bibitem[\protect\citeauthoryear{{Basant} et~al.,}{{Basant}
  et~al.}{2025}]{Basant2025}
{Basant} R.,  et~al., 2025, \mn@doi [arXiv e-prints]
  {10.48550/arXiv.2502.15074}, \href
  {https://ui.adsabs.harvard.edu/abs/2025arXiv250215074B} {p. arXiv:2502.15074}

\bibitem[\protect\citeauthoryear{{Bauer}, {Zechmeister}  \& {Reiners}}{{Bauer}
  et~al.}{2015}]{Bauer2015}
{Bauer} F.~F.,  {Zechmeister} M.,   {Reiners} A.,  2015, \mn@doi [\aap]
  {10.1051/0004-6361/201526462}, \href
  {https://ui.adsabs.harvard.edu/abs/2015A&A...581A.117B} {581, A117}

\bibitem[\protect\citeauthoryear{{Blackman}, {Szymkowiak}, {Fischer}  \&
  {Jurgenson}}{{Blackman} et~al.}{2017}]{Blackman2017}
{Blackman} R.~T.,  {Szymkowiak} A.~E.,  {Fischer} D.~A.,   {Jurgenson} C.~A.,
  2017, \mn@doi [\apj] {10.3847/1538-4357/aa5ead}, \href
  {https://ui.adsabs.harvard.edu/abs/2017ApJ...837...18B} {837, 18}

\bibitem[\protect\citeauthoryear{{Blackman}, {Ong}  \& {Fischer}}{{Blackman}
  et~al.}{2019}]{Blackman2019}
{Blackman} R.~T.,  {Ong} J.~M.~J.,   {Fischer} D.~A.,  2019, \mn@doi [\aj]
  {10.3847/1538-3881/ab24c3}, \href
  {https://ui.adsabs.harvard.edu/abs/2019AJ....158...40B} {158, 40}

\bibitem[\protect\citeauthoryear{{Blackman} et~al.,}{{Blackman}
  et~al.}{2020}]{Blackman2020}
{Blackman} R.~T.,  et~al., 2020, \mn@doi [\aj] {10.3847/1538-3881/ab811d},
  \href {https://ui.adsabs.harvard.edu/abs/2020AJ....159..238B} {159, 238}

\bibitem[\protect\citeauthoryear{{Bolton} \& {Schlegel}}{{Bolton} \&
  {Schlegel}}{2010}]{Bolton2010}
{Bolton} A.~S.,  {Schlegel} D.~J.,  2010, \mn@doi [\pasp] {10.1086/651008},
  \href {https://ui.adsabs.harvard.edu/abs/2010PASP..122..248B} {122, 248}

\bibitem[\protect\citeauthoryear{{Bouchy}, {Pepe}  \& {Queloz}}{{Bouchy}
  et~al.}{2001}]{Bouchy2001}
{Bouchy} F.,  {Pepe} F.,   {Queloz} D.,  2001, \mn@doi [\aap]
  {10.1051/0004-6361:20010730}, \href
  {https://ui.adsabs.harvard.edu/abs/2001A&A...374..733B} {374, 733}

\bibitem[\protect\citeauthoryear{{Bouchy}, {Isambert}, {Lovis}, {Boisse},
  {Figueira}, {H{\'e}brard}  \& {Pepe}}{{Bouchy} et~al.}{2009}]{Bouchy2009}
{Bouchy} F.,  {Isambert} J.,  {Lovis} C.,  {Boisse} I.,  {Figueira} P.,
  {H{\'e}brard} G.,   {Pepe} F.,  2009, in {Kern} P.,  ed.,  EAS Publications
  Series Vol. 37, EAS Publications Series. pp 247--253,
  \mn@doi{10.1051/eas/0937031}

\bibitem[\protect\citeauthoryear{{Butler}, {Marcy}, {Williams}, {McCarthy},
  {Dosanjh}  \& {Vogt}}{{Butler} et~al.}{1996}]{Butler1996}
{Butler} R.~P.,  {Marcy} G.~W.,  {Williams} E.,  {McCarthy} C.,  {Dosanjh} P.,
   {Vogt} S.~S.,  1996, \mn@doi [\pasp] {10.1086/133755}, \href
  {https://ui.adsabs.harvard.edu/abs/1996PASP..108..500B} {108, 500}

\bibitem[\protect\citeauthoryear{{Cabral} et~al.,}{{Cabral}
  et~al.}{2010}]{Cabral2010}
{Cabral} A.,  et~al., 2010, in {Atad-Ettedgui} E.,  {Lemke} D.,  eds,  Society
  of Photo-Optical Instrumentation Engineers (SPIE) Conference Series Vol.
  7739, Modern Technologies in Space- and Ground-based Telescopes and
  Instrumentation. p. 77393W, \mn@doi{10.1117/12.856901}

\bibitem[\protect\citeauthoryear{{Cabral} et~al.,}{{Cabral}
  et~al.}{2012}]{Cabral2012}
{Cabral} A.,  et~al., 2012, in {Stepp} L.~M.,  {Gilmozzi} R.,   {Hall} H.~J.,
  eds,  Society of Photo-Optical Instrumentation Engineers (SPIE) Conference
  Series Vol. 8444, Ground-based and Airborne Telescopes IV. p. 84444F,
  \mn@doi{10.1117/12.926093}

\bibitem[\protect\citeauthoryear{{Cabral} et~al.,}{{Cabral}
  et~al.}{2014}]{Cabral2014}
{Cabral} A.,  et~al., 2014, in {Ramsay} S.~K.,  {McLean} I.~S.,   {Takami} H.,
  eds,  Society of Photo-Optical Instrumentation Engineers (SPIE) Conference
  Series Vol. 9147, Ground-based and Airborne Instrumentation for Astronomy V.
  p. 91478Q, \mn@doi{10.1117/12.2055876}

\bibitem[\protect\citeauthoryear{{Carnall}}{{Carnall}}{2017}]{Carnall2017}
{Carnall} A.~C.,  2017, \mn@doi [arXiv e-prints] {10.48550/arXiv.1705.05165},
  \href {https://ui.adsabs.harvard.edu/abs/2017arXiv170505165C} {p.
  arXiv:1705.05165}

\bibitem[\protect\citeauthoryear{{Cersullo}, {Wildi}, {Chazelas}  \&
  {Pepe}}{{Cersullo} et~al.}{2017}]{Cersullo2017}
{Cersullo} F.,  {Wildi} F.,  {Chazelas} B.,   {Pepe} F.,  2017, \mn@doi [\aap]
  {10.1051/0004-6361/201629972}, \href
  {https://ui.adsabs.harvard.edu/abs/2017A&A...601A.102C} {601, A102}

\bibitem[\protect\citeauthoryear{{Cersullo}, {Coffinet}, {Chazelas}, {Lovis}
  \& {Pepe}}{{Cersullo} et~al.}{2019}]{Cersullo2019}
{Cersullo} F.,  {Coffinet} A.,  {Chazelas} B.,  {Lovis} C.,   {Pepe} F.,  2019,
  \mn@doi [\aap] {10.1051/0004-6361/201833852}, \href
  {https://ui.adsabs.harvard.edu/abs/2019A&A...624A.122C} {624, A122}

\bibitem[\protect\citeauthoryear{{Chand}, {Srianand}, {Petitjean}, {Aracil},
  {Quast}  \& {Reimers}}{{Chand} et~al.}{2006}]{Chand2006}
{Chand} H.,  {Srianand} R.,  {Petitjean} P.,  {Aracil} B.,  {Quast} R.,
  {Reimers} D.,  2006, \mn@doi [\aap] {10.1051/0004-6361:20054584}, \href
  {https://ui.adsabs.harvard.edu/abs/2006A&A...451...45C} {451, 45}

\bibitem[\protect\citeauthoryear{{Chazelas}, {Pepe}  \& {Wildi}}{{Chazelas}
  et~al.}{2012}]{Chazelas2012}
{Chazelas} B.,  {Pepe} F.,   {Wildi} F.,  2012, in \procspie. p. 845013,
  \mn@doi{10.1117/12.926188}

\bibitem[\protect\citeauthoryear{{Cheng} et~al.,}{{Cheng}
  et~al.}{2024}]{Cheng2024}
{Cheng} Y.~S.,  et~al., 2024, \mn@doi [Nature Communications]
  {10.1038/s41467-024-45924-6}, \href
  {https://ui.adsabs.harvard.edu/abs/2024NatCo..15.1466C} {15, 1466}

\bibitem[\protect\citeauthoryear{{Cook} et~al.,}{{Cook}
  et~al.}{2022}]{Cook2022}
{Cook} N.~J.,  et~al., 2022, \mn@doi [\pasp] {10.1088/1538-3873/ac9e74}, \href
  {https://ui.adsabs.harvard.edu/abs/2022PASP..134k4509C} {134, 114509}

\bibitem[\protect\citeauthoryear{{Debus}, {Sch{\"a}fer}  \& {Reiners}}{{Debus}
  et~al.}{2023}]{Debus2023}
{Debus} M.,  {Sch{\"a}fer} S.,   {Reiners} A.,  2023, \mn@doi [Journal of
  Astronomical Telescopes, Instruments, and Systems]
  {10.1117/1.JATIS.9.4.045003}, \href
  {https://ui.adsabs.harvard.edu/abs/2023JATIS...9d5003D} {9, 045003}

\bibitem[\protect\citeauthoryear{{Dumusque} et~al.,}{{Dumusque}
  et~al.}{2021}]{Dumusque2021}
{Dumusque} X.,  et~al., 2021, \mn@doi [\aap] {10.1051/0004-6361/202039350},
  \href {https://ui.adsabs.harvard.edu/abs/2021A&A...648A.103D} {648, A103}

\bibitem[\protect\citeauthoryear{{Faria} et~al.,}{{Faria}
  et~al.}{2022}]{Faria2022}
{Faria} J.~P.,  et~al., 2022, \mn@doi [\aap] {10.1051/0004-6361/202142337},
  \href {https://ui.adsabs.harvard.edu/abs/2022A&A...658A.115F} {658, A115}

\bibitem[\protect\citeauthoryear{{Figueira} et~al.,}{{Figueira}
  et~al.}{2010a}]{Figueira2010a}
{Figueira} P.,  et~al., 2010a, \mn@doi [\aap] {10.1051/0004-6361/200912681},
  \href {https://ui.adsabs.harvard.edu/abs/2010A&A...511A..55F} {511, A55}

\bibitem[\protect\citeauthoryear{{Figueira}, {Pepe}, {Lovis}  \&
  {Mayor}}{{Figueira} et~al.}{2010b}]{Figueira2010b}
{Figueira} P.,  {Pepe} F.,  {Lovis} C.,   {Mayor} M.,  2010b, \mn@doi [\aap]
  {10.1051/0004-6361/201014005}, \href
  {https://ui.adsabs.harvard.edu/abs/2010A&A...515A.106F} {515, A106}

\bibitem[\protect\citeauthoryear{{Gonz{\'a}lez Hern{\'a}ndez}, {Pepe}, {Molaro}
   \& {Santos}}{{Gonz{\'a}lez Hern{\'a}ndez}
  et~al.}{2018}]{GonzalezHernandez2018}
{Gonz{\'a}lez Hern{\'a}ndez} J.~I.,  {Pepe} F.,  {Molaro} P.,   {Santos} N.~C.,
   2018, in {Deeg} H.~J.,  {Belmonte} J.~A.,  eds, , Handbook of Exoplanets.
p.~157, \mn@doi{10.1007/978-3-319-55333-7_15710.1007/978-3-319-30648-3-157-1}

\bibitem[\protect\citeauthoryear{{Gracia Temich}, {Rasilla}, {Salata}, {Pepe},
  {Avila}, {M{\'e}gevand}, {Rebolo}  \& {Riva}}{{Gracia Temich}
  et~al.}{2018}]{GraciaTemich2018b}
{Gracia Temich} F.,  {Rasilla} J.~L.,  {Salata} S.,  {Pepe} F.,  {Avila} G.,
  {M{\'e}gevand} D.,  {Rebolo} R.,   {Riva} M.,  2018, in {Navarro} R.,  {Geyl}
  R.,  eds,  Society of Photo-Optical Instrumentation Engineers (SPIE)
  Conference Series Vol. 10706, Advances in Optical and Mechanical Technologies
  for Telescopes and Instrumentation III. p. 1070628,
  \mn@doi{10.1117/12.2312032}

\bibitem[\protect\citeauthoryear{{Griest}, {Whitmore}, {Wolfe}, {Prochaska},
  {Howk}  \& {Marcy}}{{Griest} et~al.}{2010}]{Griest2010}
{Griest} K.,  {Whitmore} J.~B.,  {Wolfe} A.~M.,  {Prochaska} J.~X.,  {Howk}
  J.~C.,   {Marcy} G.~W.,  2010, \mn@doi [\apj] {10.1088/0004-637X/708/1/158},
  \href {https://ui.adsabs.harvard.edu/abs/2010ApJ...708..158G} {708, 158}

\bibitem[\protect\citeauthoryear{{Griffin}}{{Griffin}}{1967}]{Griffin1967}
{Griffin} R.~F.,  1967, \mn@doi [\apj] {10.1086/149168}, \href
  {https://ui.adsabs.harvard.edu/abs/1967ApJ...148..465G} {148, 465}

\bibitem[\protect\citeauthoryear{{Griffin} \& {Griffin}}{{Griffin} \&
  {Griffin}}{1973}]{Griffin1973}
{Griffin} R.,  {Griffin} R.,  1973, \mn@doi [\mnras] {10.1093/mnras/162.3.255},
  \href {https://ui.adsabs.harvard.edu/abs/1973MNRAS.162..255G} {162, 255}

\bibitem[\protect\citeauthoryear{{Guy} et~al.,}{{Guy} et~al.}{2023}]{Guy2023}
{Guy} J.,  et~al., 2023, \mn@doi [\aj] {10.3847/1538-3881/acb212}, \href
  {https://ui.adsabs.harvard.edu/abs/2023AJ....165..144G} {165, 144}

\bibitem[\protect\citeauthoryear{{Hirano} et~al.,}{{Hirano}
  et~al.}{2020}]{Hirano2020}
{Hirano} T.,  et~al., 2020, \mn@doi [\pasj] {10.1093/pasj/psaa085}, \href
  {https://ui.adsabs.harvard.edu/abs/2020PASJ...72...93H} {72, 93}

\bibitem[\protect\citeauthoryear{{Hobson} et~al.,}{{Hobson}
  et~al.}{2021}]{Hobson2021}
{Hobson} M.~J.,  et~al., 2021, \mn@doi [\aap] {10.1051/0004-6361/202038413},
  \href {https://ui.adsabs.harvard.edu/abs/2021A&A...648A..48H} {648, A48}

\bibitem[\protect\citeauthoryear{{Horne}}{{Horne}}{1986}]{Horne1986}
{Horne} K.,  1986, \mn@doi [\pasp] {10.1086/131801}, \href
  {https://ui.adsabs.harvard.edu/abs/1986PASP...98..609H} {98, 609}

\bibitem[\protect\citeauthoryear{Hunter}{Hunter}{2007}]{Hunter2007}
Hunter J.~D.,  2007, \mn@doi [Computing in Science \& Engineering]
  {10.1109/MCSE.2007.55}, 9, 90

\bibitem[\protect\citeauthoryear{{Hunter} \& {Ramsey}}{{Hunter} \&
  {Ramsey}}{1992}]{Hunter1992}
{Hunter} T.~R.,  {Ramsey} L.~W.,  1992, \mn@doi [\pasp] {10.1086/133115}, \href
  {https://ui.adsabs.harvard.edu/abs/1992PASP..104.1244H} {104, 1244}

\bibitem[\protect\citeauthoryear{{Kerber}, {Nave}, {Sansonetti}, {Bristow}  \&
  {Rosa}}{{Kerber} et~al.}{2007}]{Kerber2007}
{Kerber} F.,  {Nave} G.,  {Sansonetti} C.~J.,  {Bristow} P.,   {Rosa} M.~R.,
  2007, in {Sterken} C.,  ed.,  Astronomical Society of the Pacific Conference
  Series Vol. 364, The Future of Photometric, Spectrophotometric and
  Polarimetric Standardization. p.~461

\bibitem[\protect\citeauthoryear{{King}, {Webb}, {Murphy}, {Flambaum},
  {Carswell}, {Bainbridge}, {Wilczynska}  \& {Koch}}{{King}
  et~al.}{2012}]{King2012}
{King} J.~A.,  {Webb} J.~K.,  {Murphy} M.~T.,  {Flambaum} V.~V.,  {Carswell}
  R.~F.,  {Bainbridge} M.~B.,  {Wilczynska} M.~R.,   {Koch} F.~E.,  2012,
  \mn@doi [\mnras] {10.1111/j.1365-2966.2012.20852.x}, \href
  {https://ui.adsabs.harvard.edu/abs/2012MNRAS.422.3370K} {422, 3370}

\bibitem[\protect\citeauthoryear{{Kn{\"o}ckel}, {Bodermann}  \&
  {Tiemann}}{{Kn{\"o}ckel} et~al.}{2004}]{Knöckel2004}
{Kn{\"o}ckel} H.,  {Bodermann} B.,   {Tiemann} E.,  2004, \mn@doi [European
  Physical Journal D] {10.1140/epjd/e2003-00313-4}, \href
  {https://ui.adsabs.harvard.edu/abs/2004EPJD...28..199K} {28, 199}

\bibitem[\protect\citeauthoryear{{Kokubo}, {Mori}, {Kurokawa}, {Kashiwagi},
  {Tanaka}, {Kotani}, {Nishikawa}  \& {Tamura}}{{Kokubo}
  et~al.}{2016}]{Kokubo2016}
{Kokubo} T.,  {Mori} T.,  {Kurokawa} T.,  {Kashiwagi} K.,  {Tanaka} Y.,
  {Kotani} T.,  {Nishikawa} J.,   {Tamura} M.,  2016, in {Navarro} R.,  {Burge}
  J.~H.,  eds,  Society of Photo-Optical Instrumentation Engineers (SPIE)
  Conference Series Vol. 9912, Advances in Optical and Mechanical Technologies
  for Telescopes and Instrumentation II. p. 99121R, \mn@doi{10.1117/12.2232221}

\bibitem[\protect\citeauthoryear{{Kotu{\v{s}}}, {Murphy}  \&
  {Carswell}}{{Kotu{\v{s}}} et~al.}{2017}]{Kotus2017}
{Kotu{\v{s}}} S.~M.,  {Murphy} M.~T.,   {Carswell} R.~F.,  2017, \mn@doi
  [\mnras] {10.1093/mnras/stw2543}, \href
  {https://ui.adsabs.harvard.edu/abs/2017MNRAS.464.3679K} {464, 3679}

\bibitem[\protect\citeauthoryear{{Kreider} et~al.,}{{Kreider}
  et~al.}{2022}]{Kreider2022}
{Kreider} M.~K.,  et~al., 2022, \mn@doi [arXiv e-prints]
  {10.48550/arXiv.2210.10988}, \href
  {https://ui.adsabs.harvard.edu/abs/2022arXiv221010988K} {p. arXiv:2210.10988}

\bibitem[\protect\citeauthoryear{{Landoni}, {Riva}, {Pepe}, {Conconi}, {Zerbi},
  {Cabral}, {Cristiani}  \& {Megevand}}{{Landoni} et~al.}{2014}]{Landoni2014}
{Landoni} M.,  {Riva} M.,  {Pepe} F.,  {Conconi} P.,  {Zerbi} F.~M.,  {Cabral}
  A.,  {Cristiani} S.,   {Megevand} D.,  2014, in {Ramsay} S.~K.,  {McLean}
  I.~S.,   {Takami} H.,  eds,  Society of Photo-Optical Instrumentation
  Engineers (SPIE) Conference Series Vol. 9147, Ground-based and Airborne
  Instrumentation for Astronomy V. p. 91478K, \mn@doi{10.1117/12.2056406}

\bibitem[\protect\citeauthoryear{{Liske} et~al.,}{{Liske}
  et~al.}{2008}]{Liske2008}
{Liske} J.,  et~al., 2008, \mn@doi [\mnras] {10.1111/j.1365-2966.2008.13090.x},
  \href {https://ui.adsabs.harvard.edu/abs/2008MNRAS.386.1192L} {386, 1192}

\bibitem[\protect\citeauthoryear{{Ludwig} et~al.,}{{Ludwig}
  et~al.}{2024}]{Ludwig2024}
{Ludwig} M.,  et~al., 2024, \mn@doi [Nature Communications]
  {10.1038/s41467-024-51560-x}, \href
  {https://ui.adsabs.harvard.edu/abs/2024NatCo..15.7614L} {15, 7614}

\bibitem[\protect\citeauthoryear{{Mahadevan} \& {Ge}}{{Mahadevan} \&
  {Ge}}{2009}]{Mahadevan2009}
{Mahadevan} S.,  {Ge} J.,  2009, \mn@doi [\apj] {10.1088/0004-637X/692/2/1590},
  \href {https://ui.adsabs.harvard.edu/abs/2009ApJ...692.1590M} {692, 1590}

\bibitem[\protect\citeauthoryear{{Marconi} et~al.,}{{Marconi}
  et~al.}{2022}]{Marconi2022}
{Marconi} A.,  et~al., 2022, in {Evans} C.~J.,  {Bryant} J.~J.,   {Motohara}
  K.,  eds,  Society of Photo-Optical Instrumentation Engineers (SPIE)
  Conference Series Vol. 12184, Ground-based and Airborne Instrumentation for
  Astronomy IX. p. 1218424, \mn@doi{10.1117/12.2628689}

\bibitem[\protect\citeauthoryear{{Marconi} et~al.,}{{Marconi}
  et~al.}{2024}]{Marconi2024}
{Marconi} A.,  et~al., 2024, \mn@doi [arXiv e-prints]
  {10.48550/arXiv.2407.14601}, \href
  {https://ui.adsabs.harvard.edu/abs/2024arXiv240714601M} {p. arXiv:2407.14601}

\bibitem[\protect\citeauthoryear{{Marcy} \& {Butler}}{{Marcy} \&
  {Butler}}{1992}]{Marcy1992}
{Marcy} G.~W.,  {Butler} R.~P.,  1992, \mn@doi [\pasp] {10.1086/132989}, \href
  {https://ui.adsabs.harvard.edu/abs/1992PASP..104..270M} {104, 270}

\bibitem[\protect\citeauthoryear{{Martins}}{{Martins}}{2017}]{Martins2017}
{Martins} C.~J.~A.~P.,  2017, \mn@doi [Reports on Progress in Physics]
  {10.1088/1361-6633/aa860e}, \href
  {https://ui.adsabs.harvard.edu/abs/2017RPPh...80l6902M} {80, 126902}

\bibitem[\protect\citeauthoryear{{Martins} et~al.,}{{Martins}
  et~al.}{2024}]{Martins2024}
{Martins} C.~J.~A.~P.,  et~al., 2024, \mn@doi [Experimental Astronomy]
  {10.1007/s10686-024-09928-w}, \href
  {https://ui.adsabs.harvard.edu/abs/2024ExA....57....5M} {57, 5}

\bibitem[\protect\citeauthoryear{{Mayor} \& {Queloz}}{{Mayor} \&
  {Queloz}}{1995}]{Mayor1995}
{Mayor} M.,  {Queloz} D.,  1995, \mn@doi [\nat] {10.1038/378355a0}, \href
  {https://ui.adsabs.harvard.edu/abs/1995Natur.378..355M} {378, 355}

\bibitem[\protect\citeauthoryear{{Mayor} et~al.,}{{Mayor}
  et~al.}{1983}]{Mayor1983}
{Mayor} M.,  et~al., 1983, \aaps, \href
  {https://ui.adsabs.harvard.edu/abs/1983A&AS...54..495M} {54, 495}

\bibitem[\protect\citeauthoryear{{M{\'e}gevand} et~al.,}{{M{\'e}gevand}
  et~al.}{2014}]{Megevand2014}
{M{\'e}gevand} D.,  et~al., 2014, in \procspie. p. 91471H,
  \mn@doi{10.1117/12.2055816}

\bibitem[\protect\citeauthoryear{{Metcalf}, {Fredrick}, {Terrien}, {Papp}  \&
  {Diddams}}{{Metcalf} et~al.}{2019}]{Metcalf2019a}
{Metcalf} A.~J.,  {Fredrick} C.~D.,  {Terrien} R.~C.,  {Papp} S.~B.,
  {Diddams} S.~A.,  2019, \mn@doi [Optics Letters] {10.1364/OL.44.002673},
  \href {https://ui.adsabs.harvard.edu/abs/2019OptL...44.2673M} {44, 2673}

\bibitem[\protect\citeauthoryear{{Milakovi{\'c}}, {Pasquini}, {Webb}  \& {Lo
  Curto}}{{Milakovi{\'c}} et~al.}{2020}]{Milakovic2020a}
{Milakovi{\'c}} D.,  {Pasquini} L.,  {Webb} J.~K.,   {Lo Curto} G.,  2020,
  \mn@doi [\mnras] {10.1093/mnras/staa356}, \href
  {https://ui.adsabs.harvard.edu/abs/2020MNRAS.493.3997M} {493, 3997}

\bibitem[\protect\citeauthoryear{{Milakovi{\'c}}, {Lee}, {Carswell}, {Webb},
  {Molaro}  \& {Pasquini}}{{Milakovi{\'c}} et~al.}{2021}]{Milakovic2021}
{Milakovi{\'c}} D.,  {Lee} C.-C.,  {Carswell} R.~F.,  {Webb} J.~K.,  {Molaro}
  P.,   {Pasquini} L.,  2021, \mn@doi [\mnras] {10.1093/mnras/staa3217}, \href
  {https://ui.adsabs.harvard.edu/abs/2021MNRAS.500....1M} {500, 1}

\bibitem[\protect\citeauthoryear{{Moffat}}{{Moffat}}{1969}]{Moffat1969}
{Moffat} A.~F.~J.,  1969, \aap, \href
  {https://ui.adsabs.harvard.edu/abs/1969A&A.....3..455M} {3, 455}

\bibitem[\protect\citeauthoryear{{Molaro} et~al.,}{{Molaro}
  et~al.}{2013}]{Molaro2013a}
{Molaro} P.,  et~al., 2013, \mn@doi [\aap] {10.1051/0004-6361/201321351}, \href
  {https://ui.adsabs.harvard.edu/abs/2013A&A...555A..68M} {555, A68}

\bibitem[\protect\citeauthoryear{{Murphy}, {Webb}  \& {Flambaum}}{{Murphy}
  et~al.}{2003}]{Murphy2003}
{Murphy} M.~T.,  {Webb} J.~K.,   {Flambaum} V.~V.,  2003, \mn@doi [\mnras]
  {10.1046/j.1365-8711.2003.06970.x}, \href
  {https://ui.adsabs.harvard.edu/abs/2003MNRAS.345..609M} {345, 609}

\bibitem[\protect\citeauthoryear{{Murphy} et~al.,}{{Murphy}
  et~al.}{2022}]{Murphy2021}
{Murphy} M.~T.,  et~al., 2022, \mn@doi [\aap] {10.1051/0004-6361/202142257},
  \href {https://ui.adsabs.harvard.edu/abs/2022A&A...658A.123M} {658, A123}

\bibitem[\protect\citeauthoryear{{Nave}, {Kerber}, {Den Hartog}  \& {Lo
  Curto}}{{Nave} et~al.}{2018}]{Nave2018}
{Nave} G.,  {Kerber} F.,  {Den Hartog} E.~A.,   {Lo Curto} G.,  2018, in
  \procspie. p. 1070407, \mn@doi{10.1117/12.2312286}

\bibitem[\protect\citeauthoryear{{Padovani} \& {Cirasuolo}}{{Padovani} \&
  {Cirasuolo}}{2023}]{Padovani2023}
{Padovani} P.,  {Cirasuolo} M.,  2023, \mn@doi [Contemporary Physics]
  {10.1080/00107514.2023.2266921}, \href
  {https://ui.adsabs.harvard.edu/abs/2023ConPh..64...47P} {64, 47}

\bibitem[\protect\citeauthoryear{{Pepe} et~al.,}{{Pepe}
  et~al.}{2021}]{Pepe2021}
{Pepe} F.,  et~al., 2021, \mn@doi [\aap] {10.1051/0004-6361/202038306}, \href
  {https://ui.adsabs.harvard.edu/abs/2021A&A...645A..96P} {645, A96}

\bibitem[\protect\citeauthoryear{{Perdelwitz} \& {Huke}}{{Perdelwitz} \&
  {Huke}}{2018}]{Perdelwitz2018}
{Perdelwitz} V.,  {Huke} P.,  2018, \mn@doi [\mnras] {10.1093/mnras/sty1523},
  \href {https://ui.adsabs.harvard.edu/abs/2018MNRAS.479..768P} {479, 768}

\bibitem[\protect\citeauthoryear{{Perot} \& {Fabry}}{{Perot} \&
  {Fabry}}{1899}]{Perot1899}
{Perot} A.,  {Fabry} C.,  1899, \mn@doi [\apj] {10.1086/140557}, \href
  {https://ui.adsabs.harvard.edu/abs/1899ApJ.....9...87P} {9, 87}

\bibitem[\protect\citeauthoryear{{Phillips} et~al.,}{{Phillips}
  et~al.}{2012}]{Phillips2012a}
{Phillips} D.~F.,  et~al., 2012, \mn@doi [Optics Express]
  {10.1364/OE.20.013711}, \href
  {https://ui.adsabs.harvard.edu/abs/2012OExpr..2013711P} {20, 13711}

\bibitem[\protect\citeauthoryear{{Piskunov}, {Wehrhahn}  \&
  {Marquart}}{{Piskunov} et~al.}{2021}]{Piskunov2021}
{Piskunov} N.,  {Wehrhahn} A.,   {Marquart} T.,  2021, \mn@doi [\aap]
  {10.1051/0004-6361/202038293}, \href
  {https://ui.adsabs.harvard.edu/abs/2021A&A...646A..32P} {646, A32}

\bibitem[\protect\citeauthoryear{{Planck Collaboration} et~al.,}{{Planck
  Collaboration} et~al.}{2020}]{Planck2018}
{Planck Collaboration} et~al., 2020, \mn@doi [\aap]
  {10.1051/0004-6361/201833910}, \href
  {https://ui.adsabs.harvard.edu/abs/2020A&A...641A...6P} {641, A6}

\bibitem[\protect\citeauthoryear{{Probst} et~al.,}{{Probst}
  et~al.}{2014}]{Probst2014}
{Probst} R.~A.,  et~al., 2014, {A laser frequency comb featuring sub-cm/s
  precision for routine operation on HARPS}.
p. 91471C, \mn@doi{10.1117/12.2055784}

\bibitem[\protect\citeauthoryear{{Probst} et~al.,}{{Probst}
  et~al.}{2016}]{Probst2016}
{Probst} R.~A.,  et~al., 2016, {Relative stability of two laser frequency combs
  for routine operation on HARPS and FOCES}.
p. 990864, \mn@doi{10.1117/12.2231434}

\bibitem[\protect\citeauthoryear{{Quast}, {Reimers}  \& {Levshakov}}{{Quast}
  et~al.}{2004}]{Quast2004}
{Quast} R.,  {Reimers} D.,   {Levshakov} S.~A.,  2004, \mn@doi [\aap]
  {10.1051/0004-6361:20040013}, \href
  {https://ui.adsabs.harvard.edu/abs/2004A&A...415L...7Q} {415, L7}

\bibitem[\protect\citeauthoryear{{Redman}, {Nave}  \& {Sansonetti}}{{Redman}
  et~al.}{2014}]{Redman2014}
{Redman} S.~L.,  {Nave} G.,   {Sansonetti} C.~J.,  2014, \mn@doi [\apjs]
  {10.1088/0067-0049/211/1/4}, \href
  {https://ui.adsabs.harvard.edu/abs/2014ApJS..211....4R} {211, 4}

\bibitem[\protect\citeauthoryear{{Reiners et al.}}{{Reiners et
  al.}}{2024}]{Reiners2024}
{Reiners et al.} submitted to \aap, \noop{} 2024

\bibitem[\protect\citeauthoryear{{Riva} et~al.,}{{Riva}
  et~al.}{2014}]{Riva2014b}
{Riva} M.,  et~al., 2014, in \procspie. p. 91477G, \mn@doi{10.1117/12.2056499}

\bibitem[\protect\citeauthoryear{{Ross} \& {Cardon}}{{Ross} \&
  {Cardon}}{2022a}]{Ross2022b}
{Ross} A.~J.,  {Cardon} J.~M.,  2022a, \mn@doi [Data in Brief]
  {10.1016/j.dib.2022.108038}, \href
  {https://ui.adsabs.harvard.edu/abs/2022DIB....4208038R} {42, 108038}

\bibitem[\protect\citeauthoryear{{Ross} \& {Cardon}}{{Ross} \&
  {Cardon}}{2022b}]{Ross2022a}
{Ross} A.~J.,  {Cardon} J.~M.,  2022b, \mn@doi [Journal of Molecular
  Spectroscopy] {10.1016/j.jms.2022.111589}, \href
  {https://ui.adsabs.harvard.edu/abs/2022JMoSp.38411589R} {384, 111589}

\bibitem[\protect\citeauthoryear{{Sandage}}{{Sandage}}{1962}]{Sandage1962}
{Sandage} A.,  1962, \mn@doi [\apj] {10.1086/147385}, \href
  {https://ui.adsabs.harvard.edu/abs/1962ApJ...136..319S} {136, 319}

\bibitem[\protect\citeauthoryear{{Sarmiento}, {Reiners}, {Huke}, {Bauer},
  {Guenter}, {Seemann}  \& {Wolter}}{{Sarmiento} et~al.}{2018}]{Sarmiento2018}
{Sarmiento} L.~F.,  {Reiners} A.,  {Huke} P.,  {Bauer} F.~F.,  {Guenter} E.~W.,
   {Seemann} U.,   {Wolter} U.,  2018, \mn@doi [\aap]
  {10.1051/0004-6361/201832871}, \href
  {https://ui.adsabs.harvard.edu/abs/2018A&A...618A.118S} {618, A118}

\bibitem[\protect\citeauthoryear{{Schmidt}}{{Schmidt}}{2024}]{Schmidt2024b}
{Schmidt} T.~M.,  2024, \mn@doi [\mnras] {10.1093/mnras/stae1477}, \href
  {https://ui.adsabs.harvard.edu/abs/2024MNRAS.tmp.1520S} {}

\bibitem[\protect\citeauthoryear{{Schmidt} \& {Bouchy}}{{Schmidt} \&
  {Bouchy}}{2024}]{Schmidt2024}
{Schmidt} T.~M.,  {Bouchy} F.,  2024, \mn@doi [\mnras] {10.1093/mnras/stae920},
  \href {https://ui.adsabs.harvard.edu/abs/2024MNRAS.530.1252S} {530, 1252}

\bibitem[\protect\citeauthoryear{{Schmidt} et~al.,}{{Schmidt}
  et~al.}{2021}]{Schmidt2021}
{Schmidt} T.~M.,  et~al., 2021, \mn@doi [\aap] {10.1051/0004-6361/202039345},
  \href {https://ui.adsabs.harvard.edu/abs/2021A&A...646A.144S} {646, A144}

\bibitem[\protect\citeauthoryear{{Schmidt}, {Chazelas}, {Lovis}, {Dumusque},
  {Bouchy}, {Pepe}, {Figueira}  \& {Sosnowska}}{{Schmidt}
  et~al.}{2022}]{Schmidt2022}
{Schmidt} T.~M.,  {Chazelas} B.,  {Lovis} C.,  {Dumusque} X.,  {Bouchy} F.,
  {Pepe} F.,  {Figueira} P.,   {Sosnowska} D.,  2022, \mn@doi [\aap]
  {10.1051/0004-6361/202243270}, \href
  {https://ui.adsabs.harvard.edu/abs/2022A&A...664A.191S} {664, A191}

\bibitem[\protect\citeauthoryear{{Seifahrt}, {St{\"u}rmer}, {Bean}  \&
  {Schwab}}{{Seifahrt} et~al.}{2018}]{Seifahrt2018}
{Seifahrt} A.,  {St{\"u}rmer} J.,  {Bean} J.~L.,   {Schwab} C.,  2018, in
  {Evans} C.~J.,  {Simard} L.,   {Takami} H.,  eds,  Society of Photo-Optical
  Instrumentation Engineers (SPIE) Conference Series Vol. 10702, Ground-based
  and Airborne Instrumentation for Astronomy VII. p. 107026D (\mn@eprint
  {arXiv} {1805.09276}), \mn@doi{10.1117/12.2312936}

\bibitem[\protect\citeauthoryear{{Steinmetz} et~al.,}{{Steinmetz}
  et~al.}{2008}]{Steinmetz2008}
{Steinmetz} T.,  et~al., 2008, \mn@doi [Science] {10.1126/science.1161030},
  \href {https://ui.adsabs.harvard.edu/abs/2008Sci...321.1335S} {321, 1335}

\bibitem[\protect\citeauthoryear{{St{\"u}rmer}, {Seifahrt}, {Schwab}  \&
  {Bean}}{{St{\"u}rmer} et~al.}{2017}]{Stürmer2017}
{St{\"u}rmer} J.,  {Seifahrt} A.,  {Schwab} C.,   {Bean} J.~L.,  2017, \mn@doi
  [Journal of Astronomical Telescopes, Instruments, and Systems]
  {10.1117/1.JATIS.3.2.025003}, \href
  {https://ui.adsabs.harvard.edu/abs/2017JATIS...3b5003S} {3, 025003}

\bibitem[\protect\citeauthoryear{{Terrien} et~al.,}{{Terrien}
  et~al.}{2021}]{Terrien2021}
{Terrien} R.~C.,  et~al., 2021, \mn@doi [\aj] {10.3847/1538-3881/abef68}, \href
  {https://ui.adsabs.harvard.edu/abs/2021AJ....161..252T} {161, 252}

\bibitem[\protect\citeauthoryear{{Tronsgaard}, {Buchhave}, {Wright}, {Eastman}
  \& {Blackman}}{{Tronsgaard} et~al.}{2019}]{Tronsgaard2019}
{Tronsgaard} R.,  {Buchhave} L.~A.,  {Wright} J.~T.,  {Eastman} J.~D.,
  {Blackman} R.~T.,  2019, \mn@doi [\mnras] {10.1093/mnras/stz2181}, \href
  {https://ui.adsabs.harvard.edu/abs/2019MNRAS.489.2395T} {489, 2395}

\bibitem[\protect\citeauthoryear{{Vogt} et~al.,}{{Vogt}
  et~al.}{2018}]{Vogt2018}
{Vogt} F. P.~A.,  et~al., 2018, \mn@doi [\aap] {10.1051/0004-6361/201834135},
  \href {https://ui.adsabs.harvard.edu/abs/2018A&A...618L...7V} {618, L7}

\bibitem[\protect\citeauthoryear{{Vogt} et~al.,}{{Vogt}
  et~al.}{2019}]{Vogt2019}
{Vogt} F. P.~A.,  et~al., 2019, \mn@doi [\prl]
  {10.1103/PhysRevLett.123.061101}, \href
  {https://ui.adsabs.harvard.edu/abs/2019PhRvL.123f1101V} {123, 061101}

\bibitem[\protect\citeauthoryear{{Vogt}, {Mehner}, {Figueira}, {Yu}, {Kerber},
  {Pfrommer}, {Hackenberg}  \& {Bonaccini Calia}}{{Vogt}
  et~al.}{2023}]{Vogt2023}
{Vogt} F. P.~A.,  {Mehner} A.,  {Figueira} P.,  {Yu} S.,  {Kerber} F.,
  {Pfrommer} T.,  {Hackenberg} W.,   {Bonaccini Calia} D.,  2023, \mn@doi
  [Physical Review Research] {10.1103/PhysRevResearch.5.023145}, \href
  {https://ui.adsabs.harvard.edu/abs/2023PhRvR...5b3145V} {5, 023145}

\bibitem[\protect\citeauthoryear{{Wang}, {Wright}, {MacQueen}, {Cochran},
  {Doss}, {Gibson}  \& {Schmitt}}{{Wang} et~al.}{2020}]{Wang2020}
{Wang} S.~X.,  {Wright} J.~T.,  {MacQueen} P.,  {Cochran} W.~D.,  {Doss} D.~R.,
   {Gibson} C.~A.,   {Schmitt} J.~R.,  2020, \mn@doi [\pasp]
  {10.1088/1538-3873/ab5021}, \href
  {https://ui.adsabs.harvard.edu/abs/2020PASP..132a4503W} {132, 014503}

\bibitem[\protect\citeauthoryear{{Webb}, {Flambaum}, {Churchill}, {Drinkwater}
  \& {Barrow}}{{Webb} et~al.}{1999}]{Webb1999}
{Webb} J.~K.,  {Flambaum} V.~V.,  {Churchill} C.~W.,  {Drinkwater} M.~J.,
  {Barrow} J.~D.,  1999, \mn@doi [\prl] {10.1103/PhysRevLett.82.884}, \href
  {https://ui.adsabs.harvard.edu/abs/1999PhRvL..82..884W} {82, 884}

\bibitem[\protect\citeauthoryear{{Whitmore} \& {Murphy}}{{Whitmore} \&
  {Murphy}}{2015}]{Whitmore2015}
{Whitmore} J.~B.,  {Murphy} M.~T.,  2015, \mn@doi [\mnras]
  {10.1093/mnras/stu2420}, \href
  {https://ui.adsabs.harvard.edu/abs/2015MNRAS.447..446W} {447, 446}

\bibitem[\protect\citeauthoryear{{Whitmore}, {Murphy}  \& {Griest}}{{Whitmore}
  et~al.}{2010}]{Whitmore2010}
{Whitmore} J.~B.,  {Murphy} M.~T.,   {Griest} K.,  2010, \mn@doi [\apj]
  {10.1088/0004-637X/723/1/89}, \href
  {https://ui.adsabs.harvard.edu/abs/2010ApJ...723...89W} {723, 89}

\bibitem[\protect\citeauthoryear{{Wildi}, {Pepe}, {Chazelas}, {Lo Curto}  \&
  {Lovis}}{{Wildi} et~al.}{2010}]{Wildi2010}
{Wildi} F.,  {Pepe} F.,  {Chazelas} B.,  {Lo Curto} G.,   {Lovis} C.,  2010, {A
  Fabry-Perot calibrator of the HARPS radial velocity spectrograph: performance
  report}.
p. 77354X, \mn@doi{10.1117/12.857951}

\bibitem[\protect\citeauthoryear{{Wildi}, {Pepe}, {Chazelas}, {Lo Curto}  \&
  {Lovis}}{{Wildi} et~al.}{2011}]{Wildi2011}
{Wildi} F.,  {Pepe} F.,  {Chazelas} B.,  {Lo Curto} G.,   {Lovis} C.,  2011,
  {The performance of the new Fabry-Perot calibration system of the radial
  velocity spectrograph HARPS}.
p. 81511F, \mn@doi{10.1117/12.901550}

\bibitem[\protect\citeauthoryear{{Wildi}, {Chazelas}  \& {Pepe}}{{Wildi}
  et~al.}{2012}]{Wildi2012}
{Wildi} F.,  {Chazelas} B.,   {Pepe} F.,  2012, {A passive cost-effective
  solution for the high accuracy wavelength calibration of radial velocity
  spectrographs}.
p. 84468E, \mn@doi{10.1117/12.926841}

\bibitem[\protect\citeauthoryear{{Wilken} et~al.,}{{Wilken}
  et~al.}{2010}]{Wilken2010a}
{Wilken} T.,  et~al., 2010, \mn@doi [\mnras]
  {10.1111/j.1745-3933.2010.00850.x}, \href
  {https://ui.adsabs.harvard.edu/abs/2010MNRAS.405L..16W} {405, L16}

\bibitem[\protect\citeauthoryear{{Wilken} et~al.,}{{Wilken}
  et~al.}{2012}]{Wilken2012}
{Wilken} T.,  et~al., 2012, \mn@doi [\nat] {10.1038/nature11092}, \href
  {https://ui.adsabs.harvard.edu/abs/2012Natur.485..611W} {485, 611}

\bibitem[\protect\citeauthoryear{{Wu} et~al.,}{{Wu} et~al.}{2024}]{Wu2024}
{Wu} T.-H.,  et~al., 2024, \mn@doi [Nature Photonics]
  {10.1038/s41566-023-01364-0}, \href
  {https://ui.adsabs.harvard.edu/abs/2024NaPho..18..218W} {18, 218}

\bibitem[\protect\citeauthoryear{{Zechmeister}, {Anglada-Escud{\'e}}  \&
  {Reiners}}{{Zechmeister} et~al.}{2014}]{Zechmeister2014}
{Zechmeister} M.,  {Anglada-Escud{\'e}} G.,   {Reiners} A.,  2014, \mn@doi
  [\aap] {10.1051/0004-6361/201322746}, \href
  {https://ui.adsabs.harvard.edu/abs/2014A&A...561A..59Z} {561, A59}

\bibitem[\protect\citeauthoryear{{Zhao}, {Hogg}, {Bedell}  \& {Fischer}}{{Zhao}
  et~al.}{2021a}]{Zhao2021}
{Zhao} L.~L.,  {Hogg} D.~W.,  {Bedell} M.,   {Fischer} D.~A.,  2021a, \mn@doi
  [\aj] {10.3847/1538-3881/abd105}, \href
  {https://ui.adsabs.harvard.edu/abs/2021AJ....161...80Z} {161, 80}

\bibitem[\protect\citeauthoryear{{Zhao} et~al.,}{{Zhao}
  et~al.}{2021b}]{ZhaoF2021}
{Zhao} F.,  et~al., 2021b, \mn@doi [\aap] {10.1051/0004-6361/201937370}, \href
  {https://ui.adsabs.harvard.edu/abs/2021A&A...645A..23Z} {645, A23}

\bibitem[\protect\citeauthoryear{{Zhao} et~al.,}{{Zhao}
  et~al.}{2023}]{Zhao2023}
{Zhao} L.~L.,  et~al., 2023, \mn@doi [\aj] {10.3847/1538-3881/acf83e}, \href
  {https://ui.adsabs.harvard.edu/abs/2023AJ....166..173Z} {166, 173}

\bibitem[\protect\citeauthoryear{{dell'Agostino} et~al.,}{{dell'Agostino}
  et~al.}{2014}]{dellAgostino2014}
{dell'Agostino} S.,  et~al., 2014, in {Ramsay} S.~K.,  {McLean} I.~S.,
  {Takami} H.,  eds,  Society of Photo-Optical Instrumentation Engineers (SPIE)
  Conference Series Vol. 9147, Ground-based and Airborne Instrumentation for
  Astronomy V. p. 91475W, \mn@doi{10.1117/12.2056528}

\makeatother
\end{thebibliography}




\label{lastpage}
\end{document}